\PassOptionsToPackage{svgnames}{xcolor}
\documentclass[twocolumn]{aastex62}

\usepackage{fixfoot}
\usepackage[clockwise, figuresright]{rotating}
\usepackage{amsmath}
\usepackage{color,soul}
\usepackage{multirow}
\usepackage{IEEEtrantools}
\usepackage{bm}
\usepackage{textcomp}
\usepackage{xcolor}

\defcitealias{Davis:2018}{Paper~I}

\received{2018 May 22}
\revised{2018 October 8}
\accepted{2018 October 10}
\published{2018 December 17, \href{https://iopscience.iop.org/article/10.3847/1538-4357/aae820/pdf}{\textcolor{magenta}{\apj}}, \href{http://adsabs.harvard.edu/cgi-bin/bib_query?arXiv:1810.04888}{869, 113}}

\shorttitle{The $M_{\rm BH}$--$M_{\rm *,tot}$ and $M_{\rm BH}$--$M_{\rm *,disk}$ Relations for Spiral Galaxies}
\shortauthors{Davis, Graham, and Cameron}

\begin{document}

\title{Black Hole Mass Scaling Relations for Spiral Galaxies. II. $M_{\rm BH}$--$M_{\rm *,tot}$ and $M_{\rm BH}$--$M_{\rm *,disk}$}

\correspondingauthor{Benjamin L. Davis}
\email{benjamindavis@swin.edu.au}

\author[0000-0002-4306-5950]{Benjamin L. Davis}
\affil{Centre for Astrophysics and Supercomputing, Swinburne University of Technology, Hawthorn, Victoria 3122, Australia}

\author[0000-0002-6496-9414]{Alister W. Graham}
\affil{Centre for Astrophysics and Supercomputing, Swinburne University of Technology, Hawthorn, Victoria 3122, Australia}

\author[0000-0001-8311-1491]{Ewan Cameron}
\affil{Oxford Big Data Institute, University of Oxford, Oxford OX3 7LF, United Kingdom}


\begin{abstract}

Black hole mass ($M_\text{BH}$) scaling relations are typically derived using the properties of a galaxy's bulge and samples dominated by (high-mass) early-type galaxies. 
Studying late-type galaxies should provide greater 
insight into the mutual growth of black holes and galaxies in more gas-rich
environments. 
We have used 40 spiral galaxies to establish how $M_\text{BH}$ scales with both the total stellar mass ($M_{\rm*,tot}$) 
and the disk's stellar mass, having measured the spheroid (bulge) stellar mass ($M_{\rm*,sph}$) and presented the
$M_{\rm BH}$--$M_{\rm*,sph}$ relation in \citetalias{Davis:2018}. 
The relation involving $M_{\rm*,tot}$ may be beneficial for 
estimating $M_\text{BH}$ either from pipeline data or at
higher redshift, conditions that are not ideal for the accurate isolation of the
bulge. A symmetric Bayesian analysis finds $\log\left(M_\text{BH}/M_{\sun}\right)=\left(3.05_{-0.49}^{+0.57}\right)\log\left\{M_{\rm*,tot}/[\upsilon(6.37\times10^{10}\,M_{\sun})]\right\}+(7.25_{-0.14}^{+0.13})$. The scatter from the regression of $M_\text{BH}$ on $M_{\rm*,tot}$ is 0.66\,dex; compare 0.56\,dex for $M_\text{BH}$ on $M_{\rm*,sph}$ and $0.57$\,dex for $M_\text{BH}$ on $\sigma_*$. The slope is $>2$ times that obtained using 
core-S{\'e}rsic early-type galaxies, 
echoing a similar result involving $M_{\rm*,sph}$, 
and supporting a varied growth mechanism among different morphological
types. 
This steeper relation has consequences for galaxy/black hole formation
theories, simulations, and predicting black hole masses. 
We caution that (i) an $M_\text{BH}$--$M_{\rm*,tot}$ relation built from a mixture of early-
 and late-type galaxies will find an arbitrary slope of
approximately 1--3, with no physical meaning beyond one's sample
selection, and (ii) 
evolutionary studies of the $M_\text{BH}$--$M_{\rm*,tot}$ relation need to be
mindful of the galaxy types included at each epoch. We additionally update the $M_{\rm*,tot}$--(\emph{face-on} spiral arm pitch angle) relation.


\end{abstract}

\section{Introduction}

\citet[][hereafter \citetalias{Davis:2018}]{Davis:2018} illustrate that the accurate
measurement of a galaxy's bulge (spheroid)\footnote{We shall use the terms
``spheroid'' and ``bulge'' interchangeably.} luminosity is a time-consuming task requiring a considerable level of care.  The difficulty
lies in the need to correctly decompose the surface brightness maps or
light profiles of galaxies into their constituent components, whereas
the task of just summing up all the light in a galaxy to obtain its
total luminosity is a comparatively simple process.  Nonetheless, for
some two decades astronomers have attempted this decomposition because
the centrally located supermassive black hole (SMBH) mass ($M_{\rm BH}$) is thought to
correlate with the properties of the bulge \citep{Dressler:1989}.
However, the existence of supermassive black holes in bulgeless galaxies
\citepalias[][and references therein]{Davis:2018} reveals that there is more to it than this. 

It is a small mystery why the $M_{\rm BH}$--$M_{\rm *,tot}$ (black hole mass to total galaxy stellar mass) relation has not been explored further in the
literature.  To date, its limited publication history has not been without
dramatic disagreement.  The very existence of an $M_{\rm BH}$--$M_{\rm *,tot}$
relation (or its proxy relation with bulge luminosity) has improved infinitely
from a state of nonexistence \citep{Kormendy:Gebhardt:2001} to existing, but
not being as strong a tracer of supermassive black hole  mass as the
bulge \citep{Beifiori:2012,Savorgnan:2016:II}, to being elevated to a stature
equal with that of the bulge \citep{Lasker:2014,Burcin:2018}. The latter claim would bring the $M_{\rm BH}$--$M_{\rm *,tot}$ relation in line with
suggestions that SMBH growth is a derivative of the overall potential of its
host galaxy \citep{Ferrarese:2002,Volonteri:2011}. Part of the
explanation to this small mystery undoubtedly pertains to the bend in the $M_{\rm
BH}$--$M_{\rm *,sph}$ (black hole mass to spheroid stellar mass) relation \citep{Graham:2012,Graham:Scott:2013,Scott:2013}, which steepens at the low-mass end,
departing from the near-linear relation defined by massive early-type
galaxies.  Given the
departure of these low-mass bulges from the original near-linear $M_{\rm
BH}$--$M_{\rm *,sph}$ relation, the use of total galaxy mass would have
resulted in even greater departures and perhaps the belief that black hole
mass does not correlate with galaxy mass (see \citealt{Graham:2016b} for a review
of black hole scaling relations).

The need for an $M_{\rm BH}$--$M_{\rm *,tot}$ relation becomes more critical
for nonlocal galaxies. At higher redshifts, the difficultly of accurately
separating the bulge light from the remaining light of a galaxy becomes
increasingly perilous due to the reduced spatial resolution. In the past decade, this connection has been
widely studied \citep[e.g.,][]{Merloni:2010,Bennert:2011,Cisternas:2011,Yang:2018}, with
some investigations of nonlocal galaxies going as far as to say that the $M_{\rm
BH}$--$M_{\rm *,tot}$ relation is correlated as tightly as, or tighter than, the
$M_{\rm BH}$--$M_{\rm *,sph}$
relation \citep{Peng:2007,Jahnke:2009,Bennert:2010}. In light of this, our
endeavor to focus on the $M_{\rm BH}$--$M_{\rm *,tot}$ relation in local
spiral galaxies with directly measured SMBH masses will serve as a useful
benchmark for studies of galaxies at higher redshifts, including evolutionary
studies \citep[e.g.,][]{Ivo:2003,Kollmeier:2006,Hopkins:2008,Walter:2016,Contini:2016,Burkert:2016,Yuan:2017}. This should
allow for an enrichment in our knowledge of the star formation
history \citep[e.g.,][]{Shankar:2009} and dry merger
history \citep[e.g.,][]{Jahnke:2011} of SMBH host galaxies.

The necessity for improving our knowledge of the $M_{\rm BH}$--$M_{\rm *,tot}$
relation becomes even more manifest in the lofty goals and pragmatism
surrounding large surveys of galaxies. Due to time requirements, studies of
even as few as $\approx10^2$ galaxies must rely on \emph{automated} bulge/disk
decompositions out of necessity. Even if the $M_{\rm BH}$--$M_{\rm *,sph}$
relation were intrinsically more accurate than the $M_{\rm BH}$--$M_{\rm
*,tot}$ relation, the benefits of less intrinsic scatter in the $M_{\rm
BH}$--$M_{\rm *,sph}$ relation might be overcome by the inherent measurement
errors associated with bulge/disk decompositions produced via pipeline
software. At our current technological limits, there likely exists a ceiling
in terms of survey size or redshift, beyond which the $M_{\rm BH}$--$M_{\rm
*,tot}$ relation is of greater benefit than the $M_{\rm BH}$--$M_{\rm *,sph}$
relation.

Furthermore, as discussed in \citet{Davis:2017} and \citetalias{Davis:2018}, pseudobulges have been
slandered as being pariahs and proverbial black sheep in the family of black
hole mass scaling relations. Despite one's personal opinions concerning
pseudobulges and their role in complementing/hindering studies of the $M_{\rm
BH}$--$M_{\rm *,sph}$ relation, substitution with the $M_{\rm BH}$--$M_{\rm
*,tot}$ relation allows one to seemingly escape from the stigma surrounding
pseudobulges.  Moreover, if galaxies with pseudobulges participate in the
$M_{\rm BH}$--$M_{\rm *,tot}$ relation, as they do in the $M_{\rm
BH}$--$M_{\rm *,sph}$ relation \citepalias{Davis:2018}, this may suggest that a
relation also exists with the disk stellar mass ($M_{\rm
*,disk}$). This is especially true in the case of low-mass, disk-dominated
spiral galaxies with pseudobulges as a result of the secular evolution of
their galactic
disk \citep{Combes_Sanders:1981,Combes:2009,Combes:2017}. Therefore, examining
the existence of an $M_{\rm BH}$--$M_{\rm *,disk}$ relation will be a secondary
goal of this paper, behind our primary goal of exploring the $M_{\rm
BH}$--$M_{\rm *,tot}$ relation.

Our measurement of the disk stellar masses depends on the (rather
meticulous) multicomponent galaxy decompositions presented 
in \citetalias{Davis:2018}.  In addition to modeling the disk, bulge, and bar (when present), 
rings, spiral arms, and additional nuclear components were also accounted for,
as these can otherwise bias the S\'ersic bulge parameters.

In the following section, we will briefly
recapitulate the sample selection and the light profile analysis as performed
in \citetalias{Davis:2018}, before touching on newer complements from studying the whole of
the individual galaxies. In Section~\ref{sec:compare}, we compare our galaxy
apparent magnitudes with similar studies in the literature.  In
Section~\ref{AR}, we have applied a sophisticated Bayesian analysis
to obtain the optimal $M_{\rm BH}$--$M_{\rm *,tot}$ (and $M_{\rm BH}$--$M_{\rm
*,disk}$) scaling relation for spiral galaxies, which could be highly useful,
if the scatter is acceptably low, because it does not require bulge/disk/etc.\
decompositions. We have also included the results using the
more familiar \textsc{bces} linear regression from \citet{BCES} and the
modified \textsc{fitexy} routine \citep{Press:1992,Tremaine:2002}. Finally,
given that the spiral arm pitch angle ($\phi$) traces the black hole
mass \citep{Seigar:2008,Berrier:2013,Davis:2017}, we have additionally
explored the complementary relationships between $M_{\rm *,tot}$ and $\phi$
and between $M_{\rm *,disk}$ and $\phi$, checking for consistency and insight.
We provide a discussion of our results in Section~\ref{DI} and explore how these
relations will aid in the prediction of black hole masses, particularly
intermediate-mass black holes (IMBHs). Finally, we summarize the overall
outcomes of this paper in Section~\ref{END}. In the appendices, we provide useful error propagation formulae (Appendix~\ref{App1}) and the priors and posterior values from our Bayesian regressions (Appendix~\ref{App2}).

Unless noted otherwise, all printed errors and plotted error bars represent
$1\sigma$ ($\approx 68.3\%$) confidence levels. Magnitudes are expressed in
the absolute (AB) system \citep{Oke:1974}.

\section{Data and Methodology}\label{DM}

\citet{Davis:2017} presented what we believe was, at the time, the complete sample of spiral galaxies
with directly measured SMBH masses. A contemporary analysis of astrophysical
publications had revealed 44 \textit{spiral} galaxies whose central SMBH
masses had been measured via proper motion, stellar dynamics, gaseous
dynamics, and/or astrophysical maser emission.\footnote{We have not detected
any offsets in the spiral galaxy scaling relations based on the method used to
measure the black hole mass.} This remains the largest such spiral galaxy sample
published to date, 
and references to the publications that determined the black hole masses
(listed here in Table~\ref{table:Sample} for convenience) have been 
provided in \citet{Davis:2017}. The original sample of 44 galaxies has been culled
to 40 spiral galaxies with spheroids after the removal of Cygnus~A 
(an early-type galaxy with a spiral in its intermediate-scale disk) and three
bulgeless galaxies.  Although the three bulgeless galaxies (NGC~2478, NGC~4395, and NGC~6926) could be included in our study of the $M_{\rm
BH}$--$M_{\rm *,tot}$ and $M_{\rm BH}$--$M_{\rm *,disk}$ relations, we will
use the same sample of 40 galaxies as in \citetalias{Davis:2018}, as this will enable a 
cleaner comparison of the black hole mass scaling relations for spiral
galaxies.  In particular, there is the question of how much scatter there is about
the $M_{\rm BH}$--$M_{\rm *,sph}$ relation versus the $M_{\rm BH}$--$M_{\rm
*,tot}$ relation.

Our imaging data consist primarily of \textit{Spitzer Space Telescope}
$3.6\,\micron$ imaging from the \textit{Spitzer} Survey of Stellar Structure in
Galaxies \citep[$\rm{S^4G}$;][]{Sheth:2010}, supplemented with \textit{Hubble Space Telescope} F814W and Two Micron All Sky Survey (2MASS)
$K_s$-band (2.2 $\mu$m) imaging. Isophotal fitting was performed using the software
routines \textsc{isofit} and \textsc{cmodel} \citep{Ciambur:2015}. The original images were first sky-subtracted and carefully masked for contaminating foreground and
background sources, and 
the galaxy light was then measured with a concentric set of quasi-elliptical
isophotes whose geometries were defined by their eccentric anomalies\textemdash this
allows for an 
accurate modeling of the light distribution via the inclusion of 
Fourier harmonic terms that capture deviations from pure elliptical isophotes.  The associated 
1D surface brightness profiles were then matched to models, 
which had been convolved with the 
image-dependent point spread function (PSF).

Galaxies were carefully decomposed into
multiple components, accounting for bulges, disks, bars, point sources, rings,
and spiral arms, when present, using the \textsc{profiler}
software \citep{Ciambur:2016}.  Decompositions for every galaxy can be seen in
\citetalias{Davis:2018}.  Components were identified not only based on their appearance in the
2D image (viewed at a range of contrasts) but also using the ellipticity
profile, the position angle profile, the $B_4$ Fourier harmonic profile that
captures the boxy or disky nature of the isophotes, and of course the
surface brightness profile.  Rather than adding arbitrary S\'ersic components until
some minimum $\chi^2$ value is reached\textemdash a practice seen in the literature of
late\textemdash we only include a component if we can clearly identify it with a specific
physical entity, such as a bar or a ring.
\citetalias{Davis:2018} lists which filter was used for each galaxy and shows the galaxy 
decomposition. 

\subsection{Magnitudes and Stellar Masses}\label{AbsMags}

The apparent and absolute magnitudes of the spheroids are 
listed in Table~3 of \citetalias{Davis:2018}.  Here we tabulate the total \textit{galaxy} apparent
magnitudes $(\mathfrak{m})$, determined within the \textsc{profiler}
software by integrating 
the equivalent axis\footnote{Defined by the geometric mean
$\sqrt{ab}$, where $a$ and $b$ are the major- and minor-axis lengths of a given isophote, respectively; the
``equivalent axis'' can be considered equivalent to a circle of the same
radius.} intensity model to obtain the apparent luminosity given by
\begin{equation} 
L=2\pi \int_{0}^{R_{\rm eq}>>h} I\, R_{\rm eq}\; \mathrm{d}R_{\rm eq},
\label{numerical_integration}
\end{equation}
where $I\equiv I(R_{\rm eq})$ is the intensity as a function of the
equivalent-axis radius ($R_{\rm eq}$), $h$ is the scale length of the exponential disk, and $\mathfrak{m}\propto-2.5\log{L}$. The (corrected)\footnote{We corrected for Galactic extinction, cosmological redshift dimming, and $K$-corrections, in addition to dust \citepalias[see][]{Davis:2018}.} total \textit{galaxy} absolute magnitudes $(\mathfrak{M})$ are calculated via Equation~(6) from \citetalias{Davis:2018}. 

As in \citetalias{Davis:2018}, we account for the \emph{emission} of dust at
$3.6\,\micron$ wavelengths according to the study
of \citet{Querejeta:2015}. This includes a stellar $M_*/L_*$ ratio of
$0.60\pm0.09$ from \citet{Meidt:2014} and a $\approx$25\% reduction to the
observed luminosity due to dust glow.  Our dust emission correction resulted
in $\Delta\log(M_{\rm *,tot}/M_{\sun}) = -0.12$\,dex for all of our 28
galaxies with $3.6\,\micron$ imaging.

We have applied stellar 
mass-to-light ratios (with \citealt{Chabrier:2003} initial mass
functions [IMFs]) and solar absolute magnitudes consistent with Table~1 in \citetalias{Davis:2018} 
to calculate the stellar masses. As an additional check, we calculated the stellar masses using the 2MASS magnitudes and a (stellar mass)-to-(stellar light) ratio of $0.62\pm0.08$, which yielded a very good agreement.

We derive the disk stellar mass, $M_{\rm *,disk}$, via simple subtraction such that
\begin{equation}
M_{\rm *,disk} \equiv M_{\rm *,tot} - M_{\rm *,sph}. 
\label{M_disk_def_eqn}
\end{equation}
This definition includes the spiral arms, rings, and bars (if present) as a part of the ``disk.'' Errors on $\mathfrak{m}$ are estimated from the uncertainties on the intensity model and propagated, along with uncertainties on other variables (e.g., distance), when calculating $\mathfrak{M}$ and all derivative quantities (e.g., stellar mass). For a detailed list of error propagation formulae, see Appendix~\ref{App1}. Our sample and relevant data are tabulated in Table~\ref{table:Sample}.

\begin{longrotatetable}
\begin{deluxetable*}{lllrlrlrlll}
\tabletypesize{\footnotesize}
\tablecolumns{11}
\tablecaption{Galaxy Sample and Masses\label{table:Sample}}
\tablehead{
\colhead{Galaxy Name} & \colhead{Type} & \colhead{$\lambda$} & \colhead{$A_{\lambda}$} & \colhead{$\log(M_{\rm BH}/M_{\sun})$} & \colhead{$|\phi|$} & \colhead{$\mathfrak{m}_{\lambda,{\rm tot}}$} & \colhead{$\mathfrak{M}_{\lambda,{\rm tot}}$} & \colhead{$\log(M_{\rm *,tot}/M_{\sun})$} & \colhead{$\log(M_{\rm *,disk}/M_{\sun})$} & \colhead{$B/T$} \\
\colhead{} & \colhead{} & \colhead{($\micron$)} & \colhead{(mag)} & \colhead{} & \colhead{(deg)} & \colhead{(mag)} & \colhead{(mag)} & \colhead{} & \colhead{} & \colhead{} \\
\colhead{(1)} & \colhead{(2)} & \colhead{(3)} & \colhead{(4)} & \colhead{(5)} & \colhead{(6)} & \colhead{(7)} & \colhead{(8)} & \colhead{(9)} & \colhead{(10)} & \colhead{(11)}
}
\startdata
\object{Circinus} 	&	$3.3\pm1.2$\tablenotemark{a}    &      	3.550  & 0.265          &	 $6.25^{+0.10}_{-0.12}$ 	&	 $17.0\pm3.9$ 	 	&	 $7.00\pm0.11$	&	 $-21.09\pm0.41$ 	&	 $10.62\pm0.18$ &	 $10.46\pm0.19$ 	&	 $0.31\pm0.09$	\\
\object{Cygnus~A} 	&	$5.0\pm2.0$	                &	0.8012 & 0.067 	 	&	 $9.44^{+0.11}_{-0.14}$ 	&	 $2.7\pm0.2$ 	 	&	 $12.22\pm0.43$	&	 $-25.74\pm0.43$ 	&	 $12.38\pm0.20$ &	 $11.04\pm5.33$ 	&	 $0.95\pm0.55$	\\
\object{ESO 558-G009} 	&	$3.9\pm2.1$	                &	0.8024 &  $0.610$  	&	 $7.26^{+0.03}_{-0.04}$ 	        &	 $16.5\pm1.3$ 	 	&	 $13.70\pm0.05$	&	 $-22.36\pm0.06$ 	&	 $11.03\pm0.10$ &	 $10.99\pm0.10$ 	&	 $0.07\pm0.01$	\\
\object{IC~2560} 	&	$3.4\pm0.6$\tablenotemark{a}	&	3.550  &  $0.017$  	&	 $6.49^{+0.19}_{-0.21}$ 	&	 $22.4\pm1.7$ 		&	 $11.03\pm0.08$	&	 $-21.18\pm0.92$ 	&	 $10.66\pm0.37$ &	 $10.61\pm0.37$ 	&	 $0.09\pm0.03$	\\
\object[SDSS J043703.67+245606.8]{J0437+2456}\tablenotemark{b} & \nodata\tablenotemark{a} & 0.8024 & 1.821 & $6.51^{+0.04}_{-0.05}$ 	&	 $16.9\pm4.1$ 	 	&	 $14.00\pm0.05$	&	 $-22.22\pm0.06$ 	&	 $10.97\pm0.10$ &	 $10.93\pm0.12$ 	&	 $0.09\pm0.04$	\\
\object{Milky~Way} 	&	 \nodata\tablenotemark{a}	&	0.7625 & \nodata   	&	 $6.60\pm0.02$ 	        &	 $13.1\pm0.6$ 	 	&	 \nodata	&	 $-21.25\pm0.05$\tablenotemark{c} & $10.78\pm0.10$\tablenotemark{d} &  $10.71\pm0.11$\tablenotemark{d} 	&  $0.15\pm0.02$\tablenotemark{d} \\
\object{Mrk~1029} 	&	 \nodata	                &	0.8024 &0.064 		&	 $6.33^{+0.10}_{-0.13}$ 	&	 $17.9\pm2.1$ 	 	&	 $14.47\pm0.04$	&	 $-21.44\pm0.05$ 	&	 $10.66\pm0.09$ &	 $10.57\pm0.10$ 	&	 $0.18\pm0.02$	\\
\object{NGC~0224} 	&	 $3.0\pm0.4$\tablenotemark{a}	&	3.550  &0.124  		&	 $8.15^{+0.22}_{-0.11}$ 	&	 $8.5\pm1.3$ 	 	&	 $2.45\pm0.15$\tablenotemark{e} & $-21.75\pm0.17$\tablenotemark{f} &  $10.88\pm0.10$\tablenotemark{f}  & $10.81\pm0.10$  &  $0.17\pm0.03$ \\
\object{NGC~0253} 	&	 $5.1\pm0.4$\tablenotemark{a}	&	3.550  &0.003 	 	&	 $7.00\pm0.30$ 	        &	 $13.8\pm2.3$ 	 	&	 $6.08\pm0.05$	&	 $-21.32\pm0.11$ 	&	 $10.71\pm0.08$ &	 $10.66\pm0.08$ 	&	 $0.11\pm0.01$	\\
\object{NGC~1068} 	&	 $3.0\pm0.3$\tablenotemark{a}	&	2.159  &0.010 	 	&	 $6.75\pm0.08$ 	&	 $17.3\pm1.9$ 	 	&	 $7.64\pm0.18$	&	 $-22.39\pm0.43$ 	&	 $10.78\pm0.18$ &	 $10.62\pm0.21$ 	&	 $0.31\pm0.13$	\\
\object{NGC~1097} 	&	 $3.3\pm0.5$\tablenotemark{a}   &	3.550  &0.005  	        &	 $8.38^{+0.03}_{-0.04}$ 	&	 $9.5\pm1.3$ 	 	&	 $8.65\pm0.14$	&	 $-23.05\pm0.18$ 	&	 $11.40\pm0.10$ &	 $11.27\pm0.13$ 	&	 $0.27\pm0.12$	\\
\object{NGC~1300} 	&	 $4.0\pm0.2$\tablenotemark{a}   &	3.550  &0.005  	        &	 $7.71^{+0.19}_{-0.14}$ 	&	 $12.7\pm2.0$ 	 	&	 $10.25\pm0.09$	&	 $-20.28\pm0.39$ 	&	 $10.30\pm0.17$ &	 $10.24\pm0.17$ 	&	 $0.13\pm0.06$	\\
\object{NGC~1320} 	&	 $0.9\pm0.9$	                &	3.550  &0.008 	 	&	 $6.78^{+0.24}_{-0.34}$ 	&	 $19.3\pm2.0$ 	 	&	 $11.63\pm0.10$	&	 $-20.99\pm0.97$ 	&	 $10.58\pm0.40$ &	 $10.30\pm0.41$ 	&	 $0.47\pm0.09$	\\
\object{NGC~1398} 	&	 $2.0\pm0.3$\tablenotemark{a}   &	3.550  &0.002 	 	&	 $8.03\pm0.11$ 	&	 $9.7\pm0.7$ 	 	&	 $9.03\pm0.10$	&	 $-22.65\pm0.41$ 	&	 $11.25\pm0.18$ &	 $11.14\pm0.18$ 	&	 $0.21\pm0.05$	\\
\object{NGC~2273} 	&	 $0.9\pm0.4$\tablenotemark{a}	&	0.8024 &0.107  	 	&	 $6.97\pm0.09$ 	&	 $15.2\pm3.9$ 	 	&	 $10.91\pm0.04$	&	 $-21.72\pm0.43$ 	&	 $10.77\pm0.19$ &	 $10.69\pm0.20$ 	&	 $0.16\pm0.02$	\\
\object{NGC~2748} 	&	 $4.0\pm0.1$	                &	0.8012 &0.041  	 	&	 $7.54^{+0.17}_{-0.25}$ 	&	 $6.8\pm2.2$ 	 	&	 $11.34\pm0.10$	&	 $-20.02\pm0.51$ 	&	 $10.09\pm0.22$ &	 $10.09\pm0.22$ 	        &	 \nodata	\\
\object{NGC~2960} 	&	 $0.8\pm0.9$	                &	3.550  &0.008  	 	&	 $7.06^{+0.16}_{-0.17}$ 	&	 $14.9\pm1.9$ 	 	&	 $12.34\pm0.12$	&	 $-21.68\pm0.83$ 	&	 $10.86\pm0.34$ &	 $10.65\pm0.35$ 	&	 $0.38\pm0.11$	\\
\object{NGC~2974} 	&	 \nodata\tablenotemark{a}	&	3.550  &0.010  	 	&	 $8.23^{+0.07}_{-0.08}$ 	&	 $10.5\pm2.9$ 	 	&	 $10.03\pm0.05$	&	 $-21.36\pm0.27$ 	&	 $10.73\pm0.12$ &	 $10.56\pm0.13$ 	&	 $0.32\pm0.04$	\\
\object{NGC~3031} 	&	 $2.4\pm0.6$\tablenotemark{a}   &	3.550  &0.014  		&	 $7.83^{+0.11}_{-0.07}$ 	&	 $13.4\pm2.3$ 	 	&	 $6.27\pm0.07$	&	 $-21.15\pm0.12$ 	&	 $10.65\pm0.08$ &	 $10.47\pm0.10$ 	&	 $0.33\pm0.06$	\\
\object{NGC~3079} 	&	 $6.4\pm1.1$\tablenotemark{a}	&	3.550  &0.002  	 	&	 $6.38^{+0.11}_{-0.13}$ 	&	 $20.6\pm3.8$ 	 	&	 $9.56\pm0.20$	&	 $-21.24\pm0.43$ 	&	 $10.68\pm0.18$ &	 $10.60\pm0.20$ 	&	 $0.17\pm0.08$	\\
\object{NGC~3227} 	&	 $1.5\pm0.9$\tablenotemark{a}	&	3.550  &0.004  	 	&	 $7.88^{+0.13}_{-0.14}$ 	&	 $7.7\pm1.4$ 	 	&	 $9.79\pm0.06$	&	 $-21.54\pm0.32$ 	&	 $10.80\pm0.14$ &	 $10.72\pm0.15$ 	&	 $0.17\pm0.04$	\\
\object{NGC~3368} 	&	 $2.1\pm0.7$\tablenotemark{a}	&	3.550  &0.004  	 	&	 $6.89^{+0.08}_{-0.10}$ 	&	 $14.0\pm1.4$ 	 	&	 $8.61\pm0.03$	&	 $-21.25\pm0.14$ 	&	 $10.69\pm0.09$ &	 $10.63\pm0.09$ 	&	 $0.13\pm0.02$	\\
\object{NGC~3393} 	&	 $1.2\pm0.7$\tablenotemark{a}	&	0.8024 &0.116  	 	&	 $7.49^{+0.05}_{-0.16}$ 	        &	 $13.1\pm2.5$ 	 	&	 $11.62\pm0.09$	&	 $-22.30\pm0.10$ 	&	 $11.00\pm0.10$ &	 $10.92\pm0.10$ 	&	 $0.17\pm0.03$	\\
\object{NGC~3627} 	&	 $3.1\pm0.4$\tablenotemark{a}	&	3.550  &0.006  	 	&	 $6.95\pm0.05$ 	        &	 $18.6\pm2.9$ 	 	&	 $8.35\pm0.11$	&	 $-21.48\pm0.18$ 	&	 $10.78\pm0.10$ &	 $10.73\pm0.10$ 	&	 $0.09\pm0.04$	\\
\object{NGC~4151} 	&	 $1.9\pm0.5$\tablenotemark{a}   &	3.550  &0.005  	 	&	 $7.68^{+0.15}_{-0.58}$ 	&	 $11.8\pm1.8$ 	 	&	 $10.02\pm0.10$	&	 $-21.09\pm0.31$ 	&	 $10.62\pm0.14$ &	 $10.36\pm0.16$ 	&	 $0.45\pm0.08$	\\
\object{NGC~4258} 	&	 $4.0\pm0.2$\tablenotemark{a}	&	3.550  &0.003  	 	&	 $7.60\pm0.01$ 	        &	 $13.2\pm2.5$ 	 	&	 $7.76\pm0.12$	&	 $-21.34\pm0.14$ 	&	 $10.72\pm0.09$ &	 $10.62\pm0.10$ 	&	 $0.21\pm0.09$	\\
\object{NGC~4303} 	&	 $4.0\pm0.1$\tablenotemark{a}	&	3.550  &0.004  	 	&	 $6.58^{+0.07}_{-0.26}$ 	&	 $14.7\pm0.9$ 	 	&	 $9.45\pm0.10$	&	 $-20.72\pm0.15$ 	&	 $10.48\pm0.09$ &	 $10.44\pm0.09$ 	&	 $0.09\pm0.01$	\\
\object{NGC~4388} 	&	 $2.8\pm0.7$\tablenotemark{a}   &	3.550  &0.006  	 	&	 $6.90\pm0.11$ 	&	 $18.6\pm2.6$ 	 	&	 $10.35\pm0.15$	&	 $-20.63\pm0.53$ 	&	 $10.44\pm0.22$ &	 $10.20\pm0.24$ 	&	 $0.42\pm0.08$	\\
\object{NGC~4395} 	&	 $8.8\pm0.5$\tablenotemark{a}	&	3.550  &0.003  	 	&	 $5.64^{+0.22}_{-0.12}$ 	&	 $22.7\pm3.6$ 		&	 $9.93\pm0.11$	&	 $-18.16\pm0.12$ 	&	 $9.45\pm0.08$ &	 $9.45\pm0.08$  	        &	 \nodata	\\
\object{NGC~4501} 	&	 $3.3\pm0.6$	                &	3.550  &0.007  	 	&	 $7.13\pm0.08$ 	        &	 $12.2\pm3.4$ 	 	&	 $8.76\pm0.11$	&	 $-21.21\pm0.12$ 	&	 $10.67\pm0.08$ &	 $10.53\pm0.11$ 	&	 $0.28\pm0.10$	\\
\object{NGC~4594} 	&	 $1.1\pm0.4$	                &	3.550  &0.009  	 	&	 $8.34\pm0.10$ 	        &	 $5.2\pm0.4$ 	 	&	 $7.51\pm0.28$	&	 $-22.11\pm0.31$ 	&	 $11.03\pm0.14$ &	 $10.63\pm0.40$ 	&	 $0.60\pm0.30$	\\
\object{NGC~4699} 	&	 $2.9\pm0.4$\tablenotemark{a}	&	3.550  &0.006  	 	&	 $8.34^{+0.13}_{-0.15}$ 	&	 $5.1\pm0.4$ 	 	&	 $8.84\pm0.31$	&	 $-22.75\pm0.54$ 	&	 $11.29\pm0.23$ &	 $10.79\pm0.58$ 	&	 $0.68\pm0.34$	\\
\object{NGC~4736} 	&	 $2.3\pm0.8$\tablenotemark{a}	&	3.550  &0.003  	 	&	 $6.78^{+0.09}_{-0.11}$ 	&	 $15.0\pm2.3$ 	 	&	 $7.47\pm0.08$	&	 $-20.45\pm0.11$ 	&	 $10.37\pm0.08$ &	 $10.19\pm0.09$ 	&	 $0.33\pm0.05$	\\
\object{NGC~4826} 	&	 $2.2\pm0.6$	                &	3.550  &0.007  	 	&	 $6.07^{+0.14}_{-0.16}$ 	&	 $24.3\pm1.5$ 	 	&	 $7.86\pm0.04$	&	 $-20.56\pm0.51$ 	&	 $10.41\pm0.21$ &	 $10.35\pm0.21$ 	&	 $0.14\pm0.02$	\\
\object{NGC~4945} 	&	 $6.1\pm0.6$\tablenotemark{a}   &	2.159  &0.055  	 	&	 $6.15\pm0.30$ 	        &	 $22.2\pm3.0$ 	 	&	 $6.18\pm0.13$	&	 $-21.73\pm0.17$ 	&	 $10.52\pm0.09$ &	 $10.48\pm0.09$ 	&	 $0.07\pm0.03$	\\
\object{NGC~5055} 	&	 $4.0\pm0.2$	                &	3.550  &0.003  	 	&	 $8.94^{+0.09}_{-0.11}$ 	&	 $4.1\pm0.4$ 	 	&	 $7.89\pm0.14$	&	 $-21.55\pm0.18$ 	&	 $10.81\pm0.10$ &	 $10.52\pm0.15$ 	&	 $0.48\pm0.10$	\\
\object{NGC~5495} 	&	 $5.0\pm0.4$\tablenotemark{a}	&	0.8024 &0.089  	 	&	 $7.04^{+0.08}_{-0.09}$ 	&	 $13.3\pm1.4$ 	 	&	 $12.15\pm0.07$	&	 $-23.08\pm0.07$ 	&	 $11.31\pm0.10$ &	 $11.23\pm0.10$ 	&	 $0.17\pm0.03$	\\
\object{NGC~5765b} 	&	 $2.8\pm1.5$\tablenotemark{a}	&	0.8024 &0.057  	 	&	 $7.72\pm0.05$ 	&	 $13.5\pm3.9$ 	 	&	 $13.26\pm0.04$	&	 $-22.57\pm0.19$ 	&	 $11.11\pm0.12$ &	 $11.07\pm0.12$ 	&	 $0.08\pm0.01$	\\
\object{NGC~6264} 	&	 $2.7\pm1.3$\tablenotemark{a}	&	0.8024 &0.100  	 	&	 $7.51\pm0.06$ 	        &	 $7.5\pm2.7$ 	 	&	 $13.79\pm0.03$	&	 $-22.45\pm0.27$ 	&	 $11.06\pm0.14$ &	 $11.02\pm0.14$ 	&	 $0.09\pm0.01$	\\
\object{NGC~6323} 	&	 $2.0\pm0.3$\tablenotemark{a}	&	0.8024 &0.026  	 	&	 $7.02^{+0.13}_{-0.14}$ 	&	 $11.2\pm1.3$ 		&	 $13.11\pm0.04$	&	 $-22.40\pm0.67$ 	&	 $11.04\pm0.28$ &	 $11.01\pm0.28$  	&	 $0.07\pm0.02$	\\
\object{NGC~6926} 	&	 $5.6\pm2.3$\tablenotemark{a}   &	3.550  &0.029  	 	&	 $7.74^{+0.26}_{-0.74}$ 	&	 $9.1\pm0.7$ 	 	&	 $11.71\pm0.07$	&	 $-22.80\pm0.12$ 	&	 $11.31\pm0.08$ &	 $11.31\pm0.08$ 	        &	 \nodata \\
\object{NGC~7582} 	&	 $2.1\pm0.5$\tablenotemark{a}	&	3.550  &0.002  	 	&	 $7.67^{+0.09}_{-0.08}$ 	&	 $10.9\pm1.6$ 	 	&	 $9.74\pm0.18$	&	 $-21.47\pm0.21$ 	&	 $10.77\pm0.11$ &	 $10.65\pm0.14$ 	&	 $0.24\pm0.11$	\\
\object{UGC~3789} 	&	 $1.6\pm0.6$\tablenotemark{a}	&	0.8024 &0.100  	 	&	 $7.06\pm0.05$ 	        &	 $10.4\pm1.9$ 	 	&	 $12.00\pm0.06$	&	 $-21.64\pm0.23$ 	&	 $10.74\pm0.13$ &	 $10.60\pm0.13$ 	&	 $0.28\pm0.04$	\\
\object{UGC~6093} 	&	 $3.7\pm0.8$\tablenotemark{a}	&	0.8024 &0.041  	 	&	 $7.41^{+0.04}_{-0.03}$\tablenotemark{g} &  $10.2\pm0.9$ 	&	 $13.23\pm0.08$	&	 $-22.94\pm0.17$ 	&	 $11.26\pm0.11$ &	 $11.20\pm0.12$ 	&	 $0.12\pm0.02$	\\
\enddata
\tablecomments{
Column (1): galaxy name.
Column (2): numerical morphological type from HyperLeda.
Column (3): filter wavelength \citepalias[see][Table~1]{Davis:2018}.
Column (4): Galactic extinction (in mag) due to dust attenuation in the Milky Way, at the reference wavelength listed in Column~(3), from \citet{Schlafly:Finkbeiner:2011}.
Column (5): black hole mass listed in \citet{Davis:2017}, compiled from references therein.
Column (6): logarithmic spiral arm pitch angle (\emph{face-on}, absolute value in degrees) from \citet{Davis:2017}.
Column (7): galaxy apparent magnitude (in AB mag) for the wavelength listed in Column~(3) \citepalias[calculated via][Equations (4) and (5)]{Davis:2018}.
Column (8): fully corrected galaxy absolute magnitude (in AB mag) for the wavelength listed in Column~(3) \citepalias[calculated via][Equation (6)]{Davis:2018}; \textit{Spitzer} images are additionally corrected for dust emission.
Column (9): total galaxy stellar mass (from the galaxy absolute magnitude in Column~(8), converted to a stellar mass using the appropriate solar absolute magnitude and stellar mass-to-light ratios from \citetalias{Davis:2018}, Table 1).
Column (10): disk stellar mass (via Equation~(\ref{M_disk_def_eqn})).
Column (11): bulge-to-total flux ratio.
}
\tablenotetext{a}{Indicates a barred morphology.}
\tablenotetext{b}{SDSS J043703.67+245606.8}
\tablenotetext{c}{From \citet{Okamoto:2013}.} 
\tablenotetext{d}{From \citet{Licquia:Newman:2015}.}
\tablenotetext{e}{From \citet{Savorgnan:2016}.}
\tablenotetext{f}{From \citet{Savorgnan:2016:II}.}
\tablenotetext{g}{From \citet{Zhao:2018}.}
\end{deluxetable*}
\end{longrotatetable}

\subsection{Colors}

\begin{figure} 
\includegraphics[clip=true,trim= 0mm 0mm 0mm 0mm,width=\columnwidth]{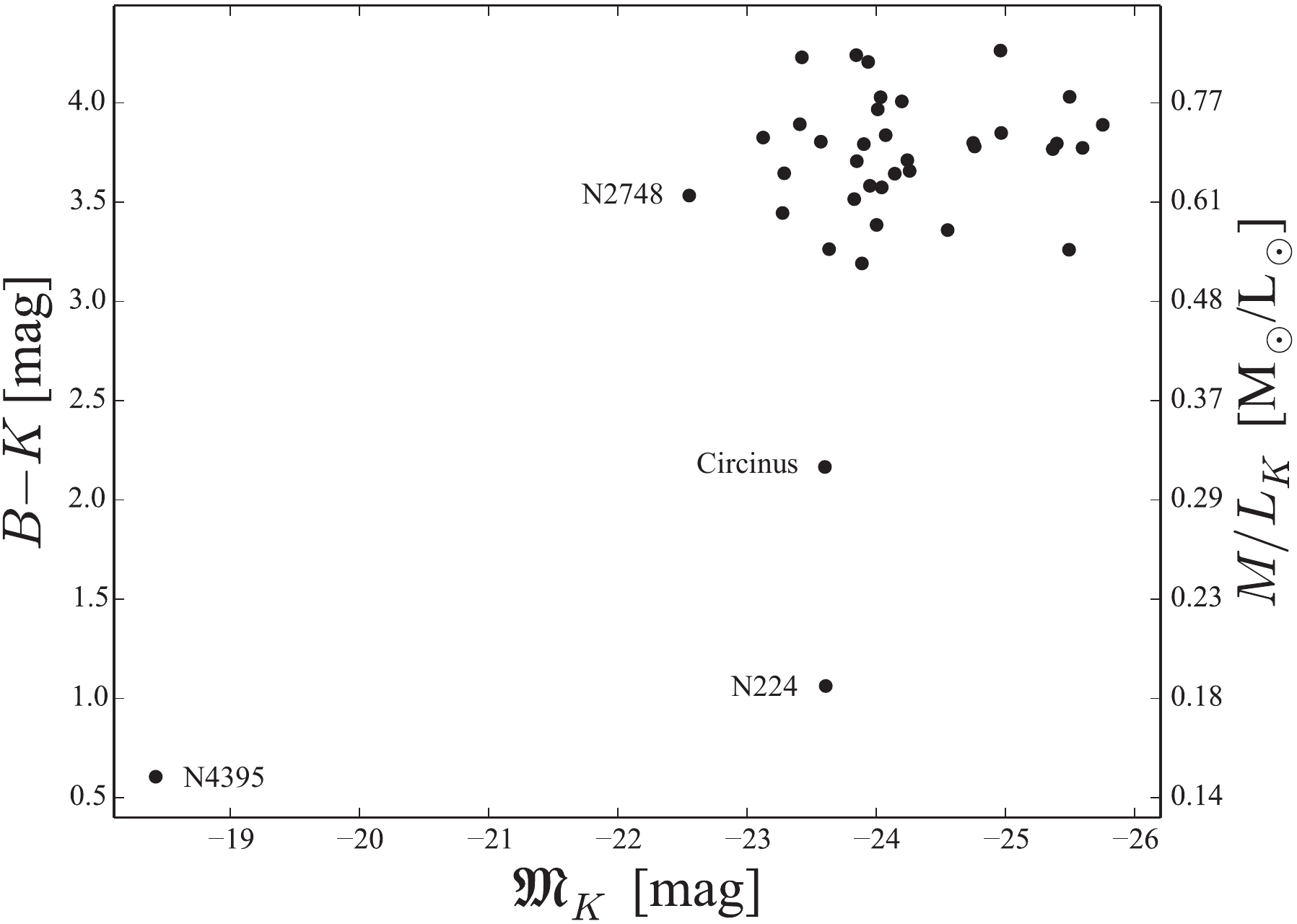} 
\caption{$B-K$ color-magnitude diagram for our spiral galaxy sample. The associated $K$-band stellar mass-to-light ratios \citep[via the prescription in][based on the $B-K$ color]{Bell:2001} are shown on the right axis.}
\label{fig:colors} 
\end{figure}

Our sample represents all of the currently known spiral galaxies with
directly measured black hole masses. However, the colors of these
spiral galaxies are not representative of the full spiral galaxy
population.  As can be seen in Figure~\ref{fig:colors}, the majority
of our galaxies have colors clustered around a median $B-K$
color equal to $3.77\pm0.22\,{\rm mag}$, where the $B$-band magnitudes have come
from the Third Reference Catalog of Bright Galaxies 
\citep[RC3;][]{RC3} and the $K$-band magnitudes have come from
2MASS.\footnote{\url{http://www.ipac.caltech.edu/2mass}, \citet{Jarrett:2000}.}
Furthermore, we have corrected the magnitudes for Galactic extinction \citep{Schlafly:Finkbeiner:2011}.
This galaxy selection ``bias'' is not unexpected though: given the
necessity to resolve the gravitational sphere of influence around the
black holes, only the most massive black holes can be directly
measured, yielding host spiral galaxies that are more massive and
redder than a general population of ``blue cloud'' spiral galaxies \citep{Cassata:2007}. 
The roughly constant color gives additional support to our use of a
constant stellar mass-to-light ratio in the \textit{Spitzer} $3.6\,\micron$ band. 
That is, the lack of a trend between color and magnitude in our 
sample suggests that our galaxies' stellar masses should not simply be
thought of as scaled luminosities, but indeed as stellar masses.  

While red spiral galaxies are known to have a range of morphologies
\citep{Masters:2010,Chilingarian:2012}, they are rare at stellar
masses less than $10^{10}\,M_{\sun}$. One may speculate whether our spiral
galaxies are red because they have black holes that are massive
enough to have blown out their gas and quench their star formation. 
Arguably, \citet{Savorgnan:2016:II} may, therefore, have prematurely
referred to the spiral galaxy sequence in the $M_{\rm BH}$--$M_{\rm
*,sph}$ diagram as a blue sequence. However, it is known that some
low-mass, blue, spiral galaxies possess active galactic nuclei --- for
example, NGC~4395 \citep{Brok:2015} and LEDA~87300
\citep{Baldassare:2015,Graham:2016}---and therefore, we are simply
probing the red end of the blue sequence. LEDA~87300 has a $g^{\prime}-r^{\prime}$
color equal to $0.41\,{\rm mag}$ \citep{Graham:2016}, which is slightly bluer than
NGC~4395 with $g^{\prime}-r^{\prime}=0.50\,{\rm mag}$.\footnote{Here, the magnitudes are obtained from the Sloan Digital Sky Survey Data Release 6 (\url{http://www.sdss.org/dr6/products/catalogs/index.html}) and subsequently corrected for Galactic extinction \citep{Schlafly:Finkbeiner:2011}.}

\vspace{2cm}

\subsection{$\upsilon$}\label{sec:upsilon}

In \citetalias{Davis:2018}, we introduced a new parameter,\footnote{The value of
$\upsilon$ has \emph{no} affect on the slope of the scaling
relations.} $\upsilon$. It is our hope that readers may
easily apply the scaling relations herein to their own studies by calibrating
to their adopted initial mass function.  This conversion is
accomplished in a fashion similar to that achieved via $h$ in
cosmological conversions.  Often, cosmologists will normalize their
cosmologies, where $h=1$ implies a Hubble constant of $100\,\text{km\,s}^{-1}\,\text{Mpc}^{-1}$. Similarly, researchers who conduct simulations of
galaxies will often normalize their initial-mass-function-dependent
stellar mass-to-light ratio, $\Upsilon_*$. 

For example, from 40 of our 43 galaxies with available
photometry on NED,\footnote{\url{http://nedwww.ipac.caltech.edu}} we
find that $\upsilon=1.08\pm0.15$ when comparing our galaxy
stellar masses (Table~\ref{table:Sample}) to those predicted using 2MASS $K$-band
magnitudes and the $B-K$ color-dependent stellar mass-to-light
ratios from \citet{Bell:2001}. Alternatively, to adjust our stellar
masses to match those predicted from the Sloan Digital Sky Survey (SDSS) $i^\prime$-band magnitudes and
$g^{\prime}-i^{\prime}$ color-dependent stellar mass-to-light ratios
from \citet{Bell:2003}, \citet{Taylor:2011}, or \citet{Roediger:2015}
would require $\upsilon=0.81\pm0.17$, $0.36\pm0.09$, or
$0.51\pm0.10$, respectively.

\vspace{1.4cm}

\section{Comparison of $3.6\,\micron$ Magnitudes}\label{sec:compare}

\subsection{\citet{Savorgnan:2016}}\label{Sav_compare}

\begin{figure}
\includegraphics[clip=true,trim= 0mm 0mm 0mm 0mm,width=\columnwidth]{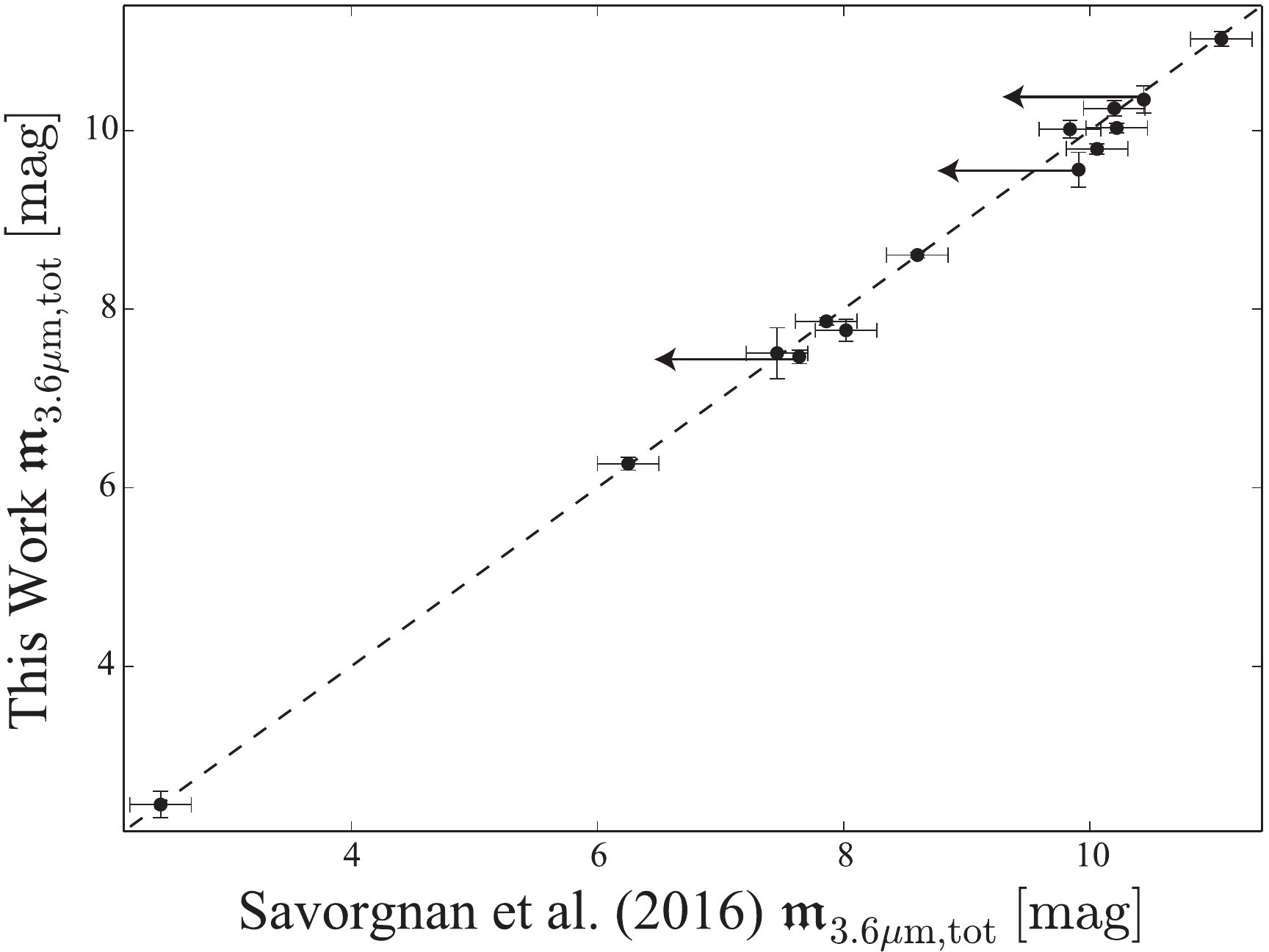}
\caption{Comparison of the total $3.6\,\micron$ apparent magnitudes (with a
1:1 dashed line) from 14 spiral galaxies in common with \citet{Savorgnan:2016}
yields $\Delta_{rms,\perp}=0.07$\,mag. Three of the values
from \citet{Savorgnan:2016} are upper limits, indicated with
arrows. Note that the Vega magnitudes from \citet{Savorgnan:2016} have been
converted here to the AB magnitude system.}
\label{SG16_compare_plot}
\end{figure}

We first compare our data set with that of \citet{Savorgnan:2016}. Our work builds on those studies by analyzing
many of the same galaxies, in the same $3.6\,\micron$ passband, and with
similar decompositional methodology. Figure \ref{SG16_compare_plot} shows that
our total apparent magnitudes match well with \citet{Savorgnan:2016:II}, with
an rms scatter
$\Delta_{rms,\perp}=0.07$\,mag.\footnote{Throughout \citetalias{Davis:2018} and this work, we
analyze the agreement (in diagrams with the same quantity on both axes) by
calculating the orthogonal rms scatter ($\Delta_{rms,\perp}$)
about the 1:1 line, with $\Delta_{rms,\perp} = \Delta_{\rm rms}/\sqrt{2}.$} We
find this high level of agreement to be four times tighter than between the
spheroid apparent magnitudes for the same galaxies, reflective of the
challenges in obtaining bulge magnitudes.

The multicomponent surface brightness profile decompositional methodology
of \citet{Savorgnan:2016} largely agrees with ours. Both methods involve
decomposition of 1D surface brightness profiles and do not use a
signal-to-noise weighting scheme as a result of the propensity for things to go awry
at the centers of galaxies and consequentially wreak havoc on the fit. 
Differing from \citet{Savorgnan:2016}, we have used the software packages from \citet{Ciambur:2015,Ciambur:2016},
which allowed us to better model the quasi-elliptical shape of the isophotes
and perform more realistic PSF convolutions with our models.

\subsection{Spitzer Survey of Stellar Structure in Galaxies}

\begin{figure}
\includegraphics[clip=true,trim= 9mm 5mm 20mm 15mm,width=\columnwidth]{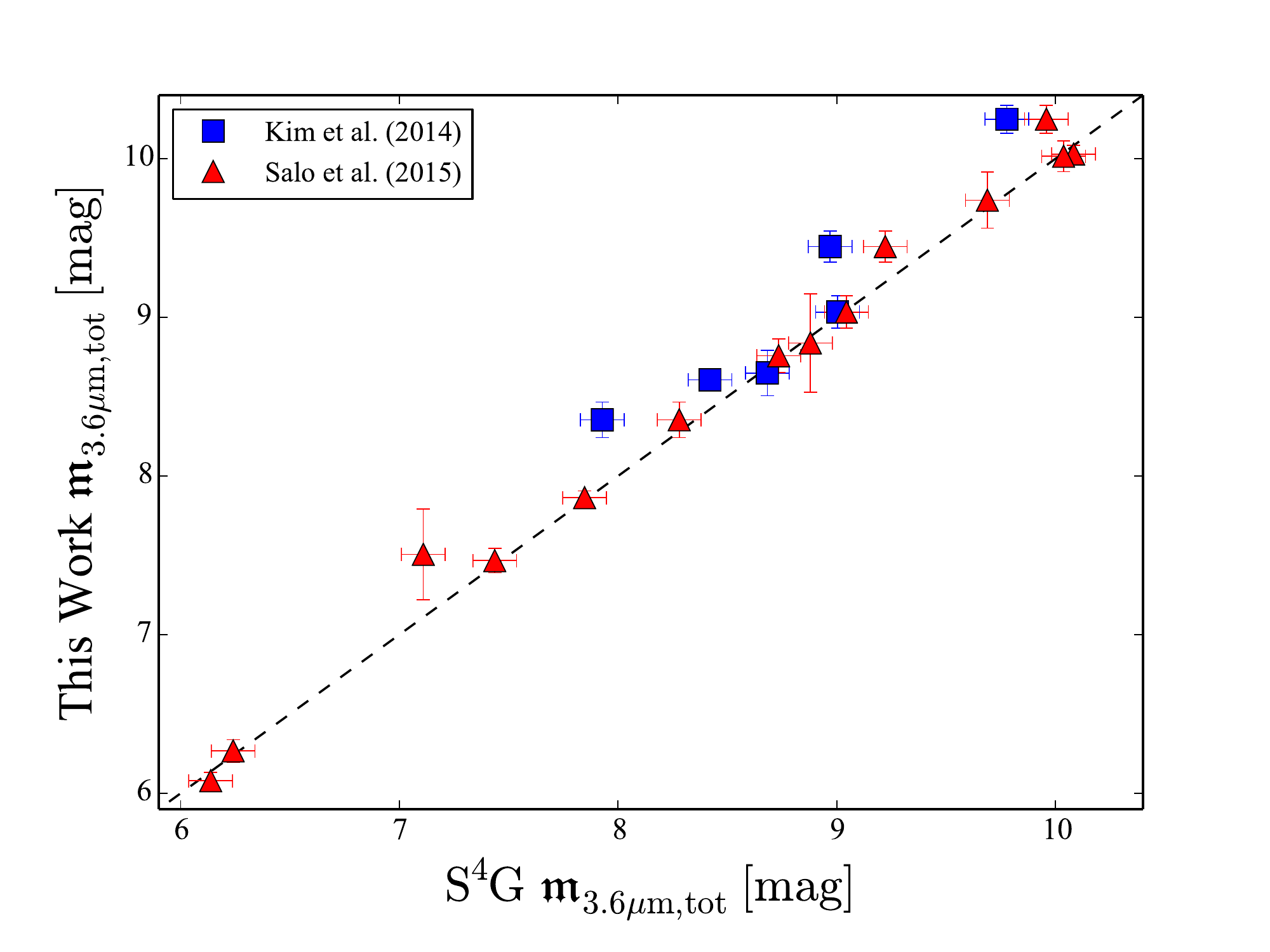}
\caption{Comparison of the total $3.6\,\micron$ apparent magnitudes (with 1:1
dashed line) for data from 14 spiral galaxies that are in common with the
S$^4$G sample from \citet{Salo:2015} plus six from \citet{Kim:2014}. The
agreement is such that $\Delta_{rms,\perp}=0.09$\,mag \citep{Kim:2014} and
$\Delta_{rms,\perp}=0.06$\,mag \citep{Salo:2015}. Note that the S$^4$G does
not provide error estimates, so we have added error bars equivalent to our
median error.} 
\label{S4G_compare}
\end{figure}

We have additionally compared our total apparent magnitudes to those from
the \textit{Spitzer} Survey of Stellar Structure in
Galaxies \citep[S$^4$G:][]{Kim:2014,Salo:2015}, which also examines an overlapping
set of galaxies with our sample, and with identical imaging. In
Figure~\ref{S4G_compare}, we find a low level of scatter of
$\Delta_{rms,\perp}=0.09$\,mag with the six common galaxies from \citet{Kim:2014}
and $\Delta_{rms,\perp}=0.06$\,mag with the 14 common galaxies
from \citet{Salo:2015}. These low levels of scatter are approximately one-half and one-fifth,
respectively, of the scatter found among the spheroid apparent magnitudes
for these same galaxies \citepalias[see][]{Davis:2018}, and it is similar to the scatter found above
from \citet{Savorgnan:2016}.

Although we analyzed identical \textit{Spitzer} images to the
S$^4$G, they performed a 2D (opposed to our 1D) decomposition of the galaxies' observed surface brightness distributions. \citet{Kim:2014} and
\citet{Salo:2015} utilized the \textsc{budda}
\citep{BUDDA:2004,Gadotti:2008,Gadotti:2009} and \textsc{galfit}
\citep{Peng:2002,Peng:2010} software routines, respectively. As
\citet{Ciambur:2016} points out, pros and cons are associated with both
1D and 2D decomposition techniques. Neither approach is perfect, mainly as a result of
some form of azimuthal averaging.

In particular, 1D codes work on
azimuthally averaged isophotes, which collectively capture the radial
gradients of the Fourier harmonic terms in these isophotes.  Indeed, the
discovery and measurement of the isophotal $B_6$ Fourier harmonic, as well as its
association with (peanut shell)-shaped bulges, were made via 1D image
analysis using \textsc{isofit} \citep{Ciambur:2015, Ciambur_Graham:2016}. One of the advantages with collapsing a 2D image into a set of 1D profiles (e.g.,
surface brightness, ellipticity, position angle, and Fourier terms) is that the
fitted galaxy model components, in one's subsequent decomposition of the light
profile, account for these variations.  That is, for example, one is not
trying to fit a triaxial bulge with a 2D model that has a constant
position angle and ellipticity, but rather one accounts for these
isophotal twists and changes with radius.  A fuller discussion 
can be found in \citet{Ciambur:2015,Ciambur:2016}. 

\section{Regression Analyses}\label{AR}

Regression analysis in astronomy is simultaneously a crucial but inherently
difficult task. Astronomical data are plagued with many complicating
conditions arising from the difficulty of collecting data from great
distances, selection effects, heteroscedasticity,
etc. This complicates one's data sets, which ultimately must be compressed down
to two numbers in a linear regression: slope and intercept. As a result,
astronomers have developed many varied statistical approaches, manifest in the
myriad of computer codes.

The astronomical community has been rapidly adopting Bayesian statistical
methods over the past couple of decades \citep[e.g.,][]{HyperFit,Pihajoki:2017}. \citet{Andreon:2013} provide a review
of measurement errors and scaling relations in astrophysics and advocate for
Bayesian regression techniques.
In deriving the $M_{\rm BH}$--$M_{\rm *,tot}$ and $M_{\rm BH}$--$M_{\rm
*,disk}$ scaling relations in this paper, our custom Bayesian analysis \citepalias[detailed in][]{Davis:2018} explores both a \textit{conditional} minimization of offsets in the
vertical $\log M_{\rm BH}$ direction about the fitted line and a \textit{symmetric} treatment of
the data in both directions.

To date, many, if not most, of the published black hole mass scaling relations have been
derived using either 
the \textsc{bces} \citep[Bivariate Correlated Errors and intrinsic
Scatter;][]{BCES} or 
the \textsc{mpfitexy} \citep{Press:1992,Tremaine:2002,Bedregal:2006,Novak:2006,Markwardt:2009,Williams:2010,MPFIT}
routine.  
For comparison, the data are additionally analyzed here using both of these
more familiar routines.  Reassuringly, when performing a ``forward'' regression
(minimizing the vertical offset of the data about the fitted line), an
``inverse'' regression (minimizing the horizontal offset of the data about the
fitted line), or instead treating the data symmetrically (here we use a line
that bisects the slopes of the above two lines), we recover consistent scaling relations using each of these methods.

Ordinary least-squares regression bisection has been recommended for
treating variables symmetrically for nearly three decades since the seminal
work by \citet{Isobe:1990}. While our Bayesian analysis provides a symmetrical
treatment of the $(X,\,Y)$ data sets, as does the \citet{BCES} routine, a
symmetric treatment of the data can also be obtained when using the
asymmetrical \textsc{mpfitexy} routine by bisecting the results of the
``forward'' and ``inverse'' linear regressions \citep[see, e.g.,][]{Novak:2006}.
Although \citet{Graham:2009} used \textsc{bces}, \textsc{mpfitexy}, and a
different Bayesian code from \citet{Kelly:2007}, and found that they all
provided consistent results (see also \citealt{Park:2012}, for a more detailed
report), it remains prudent to check, especially as the \textsc{bces} routine
can struggle when the measurement errors are large \citep{Tremaine:2002}. The
recovery of slopes and intercepts that are consistent with each other will also
provide confidence that one has not been led astray by a single statistical 
analysis.

The primary sources of uncertainty on the stellar mass estimates in our
analyses consist of the individual uncertainties on the stellar mass-to-light ratios,
distances, and the photometry. 
The median relative uncertainties that we assigned to these terms in \citetalias{Davis:2018} are 15\%, 10\%, and 10\%, respectively.
 
\subsection{Relations with Black Hole Mass ($M_{\rm BH}$)}

\subsubsection{The $M_{\rm BH}$--$M_{\rm *,tot}$ Relation}\label{sec:M_BH-M_tot}

\begin{figure}
\includegraphics[clip=true,trim= 4mm 4mm 1mm 1mm,width=\columnwidth]{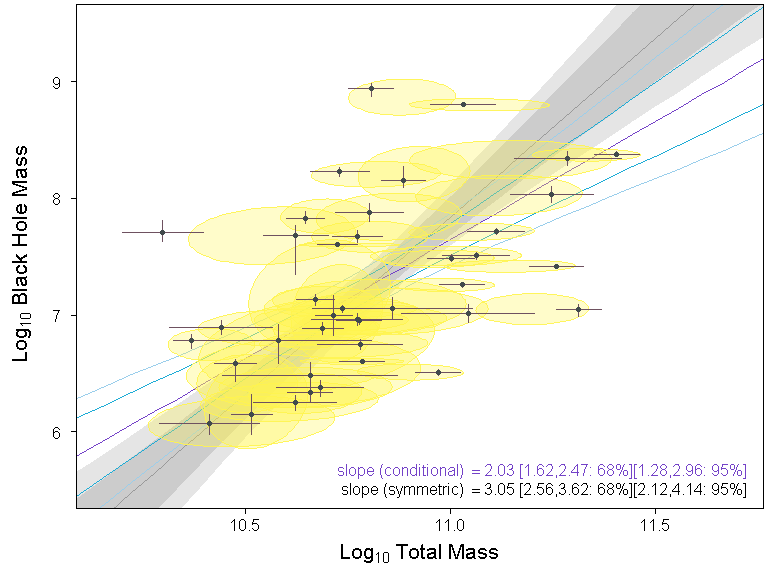}
\caption{The \textit{symmetric} (gray) Bayesian line of best fit (see Equation~(\ref{Ewan_M_BH-M_tot_eqn})) is presented as its pointwise median with $\pm$68\% and $\pm$95\% (shaded) intervals, while the $\pm$68\% posterior estimates of the true stellar total and black hole mass of each galaxy are highlighted in yellow. The \textit{conditional} (purple) line of best fit is additionally supplied with similar (cyan) error intervals. Masses are in units of solar masses.}
\label{Ewan_M_BH-M_tot_plot}
\end{figure}

Our $\left(\log M_{\rm *,tot},\, \log M_{\rm BH}\right)$ data set has a Pearson correlation
coefficient $r = 0.47$, and a $p$-value probability equal to $1.97\times10^{-3}$ that the null hypothesis is true. The Spearman rank-order
correlation coefficient $r_s=0.53$, with
$p_s=4.53\times10^{-4}$ that the null hypothesis is true. We find the data to be slightly less correlated than the $\left(\log M_{\rm
*,sph}, \log M_{\rm BH}\right)$ data set we presented in \citetalias{Davis:2018}, which had $r = 0.66$ with $p=4.49\times10^{-6}$ and $r_s=0.62$ with $p_s=2.38\times10^{-5}$. Of course, one
should bare in mind that the Pearson and Spearman correlation coefficients are
ignorant of the error bars assigned to each datapoint. As such, one should
turn to the uncertainty on the slope of the relation constructed through an
analysis that allows for these
errors. Our \textit{symmetric} Bayesian analysis yields the
following equation:
\begin{IEEEeqnarray}{rCl}
\log\left( \frac{M_{\rm BH}}{M_{\sun}}\right ) & = & \left(3.05_{-0.49}^{+0.57}\right)\log\left[ \frac{M_{\rm *,tot}}{\upsilon(6.37\times10^{10}\,M_{\sun})}\right ] \nonumber \\ 
&& +\> \left(7.25_{-0.14}^{+0.13}\right),
\label{Ewan_M_BH-M_tot_eqn}
\end{IEEEeqnarray}
with $\Delta_{\rm rms} = 0.79$\,dex and $\epsilon=0.69$\,dex in the $\log{M_{\rm BH}}$ direction (see Figure~\ref{Ewan_M_BH-M_tot_plot}). This regression, as well as all subsequent regressions in this work, is provided in Table~\ref{table:bent}. We note that the minimum vertical scatter is achieved when using the \textit{conditional} regression, which yields $\Delta_{\rm rms} = 0.66$\,dex and $\epsilon=0.61$\,dex.

\begin{deluxetable*}{lcccccccrrrr}
\tablecolumns{11}
\tabletypesize{\footnotesize}
\tablecaption{Linear Regressions\label{table:bent}}
\tablehead{
\colhead{Regression} & \colhead{Minimization} & \colhead{$\alpha$} & \colhead{$\beta$} & \colhead{$\epsilon$} & \colhead{$\Delta_{\rm rms}$} & \colhead{} & \colhead{$r$} & \colhead{$\log{p}$} & \colhead{$r_s$} & \colhead{$\log{p_s}$} \\
\colhead{} & \colhead{} & \colhead{} & \colhead{(dex)} & \colhead{(dex)} & \colhead{(dex)} & \colhead{} & \colhead{} & \colhead{(dex)} & \colhead{} & \colhead{(dex)} \\
\colhead{(1)} & \colhead{(2)} & \colhead{(3)} & \colhead{(4)} & \colhead{(5)} & \colhead{(6)} & \colhead{} & \colhead{(7)} & \colhead{(8)} & \colhead{(9)} & \colhead{(10)}
}
\startdata
\multicolumn{11}{c}{40 Late-type Galaxies with S{\'e}rsic Bulges} \\
\hline
\multicolumn{11}{c}{The $M_{\rm BH}$--$M_{\rm *,tot}$ Relation: $\log(M_{\rm BH}/M_{\sun})=\alpha\log\left\{M_{\rm *,tot}/[\upsilon(6.37\times10^{10}\,M_{\sun})]\right\}+\beta$} \\
\hline
Bayesian & \textit{Symmetric} & $3.05_{-0.49}^{+0.57}$ & $7.25_{-0.14}^{+0.13}$ & 0.69 & 0.79 & \multirow{8}{*}{$\left\}\begin{tabular}{@{}l@{}} \\ \\ \\ \\ \\ \\ \\ \\ \end{tabular}\right.$} & \multirow{8}{*}{$0.47$} & \multirow{8}{*}{$-2.71$} & \multirow{8}{*}{$0.53$} & \multirow{8}{*}{$-3.34$} \\
Bayesian & $M_{\rm BH}$ & $2.03_{-0.41}^{+0.44}$ & $7.25_{-0.14}^{+0.13}$ & 0.61 & 0.66 & & & & \\
\textsc{bces} & \textit{Symmetric} & $3.05\pm0.70$ & $7.25\pm0.13$ & 0.70 & 0.79 & & & \\
\textsc{bces} & $M_{\rm BH}$ & $2.04\pm0.73$ & $7.26\pm0.11$ & 0.61 & 0.66 & & & & \\
\textsc{bces} & $M_{\rm *,tot}$ & $5.60\pm1.57$ & $7.25\pm0.21$ & 1.11 & 1.31 & & & & \\
\textsc{mpfitexy} & \textit{Symmetric} & $2.65\pm0.65$ & $7.26\pm0.14$ & 0.65 & 0.73 & & & \\
\textsc{mpfitexy} & $M_{\rm BH}$ & $1.62\pm0.39$ & $7.27\pm0.10$ & $0.60$ & 0.64 & & & & \\
\textsc{mpfitexy} & $M_{\rm *,tot}$ & $5.94\pm1.88$ & $7.25\pm0.23$ & 1.18 & 1.39 & & & & \\
\hline
\multicolumn{11}{c}{The $M_{\rm BH}$--$M_{\rm *,disk}$ Relation: $\log(M_{\rm BH}/M_{\sun})=\alpha\log\left\{M_{\rm *,disk}/[\upsilon(4.98\times10^{10}\,M_{\sun})]\right\}+\beta$} \\
\hline
Bayesian & \textit{Symmetric} & $2.83_{-0.42}^{+0.55}$ & $7.24\pm0.13$ & 0.78 & 0.91 & \multirow{5}{*}{$\left\}\begin{tabular}{@{}l@{}} \\ \\ \\ \\ \\ \\ \\ \\ \end{tabular}\right.$} & \multirow{8}{*}{$0.28$} & \multirow{8}{*}{$-1.09$} & \multirow{8}{*}{$0.34$} & \multirow{8}{*}{$-1.51$} \\
Bayesian & $M_{\rm BH}$ & $1.74_{-0.35}^{+0.43}$ & $7.24\pm0.13$ & 0.67 & 0.75 & & & & \\
\textsc{bces} & \textit{Symmetric} & $2.72\pm1.07$ & $7.30\pm0.14$ & 0.77 & 0.88 & & & \\
\textsc{bces} & $M_{\rm BH}$ & $1.48\pm0.87$ & $7.28\pm0.12$ & 0.66 & 0.72 & & & & \\
\textsc{bces} & $M_{\rm *,disk}$ & $9.12\pm4.70$ & $7.41\pm0.42$ & 2.08 & 2.43 & & & & \\
\textsc{mpfitexy} & \textit{Symmetric} & $2.38\pm0.86$ & $7.26\pm0.17$ & 0.73 & 0.83 & & & & \\
\textsc{mpfitexy} & $M_{\rm BH}$ & $1.24\pm0.39$ & $7.26\pm0.11$ & $0.66$ & 0.70 & & & & \\
\textsc{mpfitexy} & $M_{\rm *,disk}$ & $8.53\pm4.67$ & $7.26\pm0.37$ & 1.94 & 2.28 & & & & \\
\hline
\multicolumn{11}{c}{The $M_{\rm *,tot}$--$\phi$ Relation: $\log(M_{\rm *,tot}/M_{\sun})=\alpha\left[|\phi|-13\fdg4 \right ]{\rm deg}^{-1}+\beta+\log \upsilon$} \\
\hline
\textsc{bces} & \textit{Symmetric} & $-0.053\pm0.013$ & $10.82\pm0.04$ & $0.20$ & 0.25 & \multirow{6}{*}{$\left\}\begin{tabular}{@{}l@{}} \\ \\ \\ \\ \\ \\ \end{tabular}\right.$} & \multirow{6}{*}{$-0.52$} & \multirow{6}{*}{$-3.29$} & \multirow{6}{*}{$-0.58$} & \multirow{6}{*}{$-4.04$} \\
\textsc{bces} & $M_{\rm *,tot}$ & $-0.038\pm0.008$ & $10.82\pm0.04$ & $0.19$ & 0.23 & & & & \\
\textsc{bces} & $|\phi|$ & $-0.068\pm0.024$ & $10.83\pm0.05$ & $0.23$ & 0.29 & & & & \\
\textsc{mpfitexy} & \textit{Symmetric} & $-0.061\pm0.013$ & $10.80\pm0.05$ & $0.21$ & 0.27 & & & \\
\textsc{mpfitexy} & $M_{\rm *,tot}$ & $-0.035\pm0.009$ & $10.81\pm0.04$ & 0.19 & 0.23 & & & & \\
\textsc{mpfitexy} & $|\phi|$ & $-0.087\pm0.018$ & $10.79\pm0.06$ & $0.28$ & $0.36$ & & & & \\
\hline
\multicolumn{11}{c}{The $M_{\rm *,disk}$--$\phi$ Relation: $\log(M_{\rm *,disk}/M_{\sun})=\alpha\left[|\phi|-13\fdg4 \right ]{\rm deg}^{-1}+\beta+\log \upsilon$} \\
\hline
\textsc{bces} & \textit{Symmetric} & $-0.054\pm0.022$ & $10.70\pm0.05$ & 0.24 & 0.30 & \multirow{6}{*}{$\left\}\begin{tabular}{@{}l@{}} \\ \\ \\ \\ \\ \\ \end{tabular}\right.$} & \multirow{6}{*}{$-0.35$} & \multirow{6}{*}{$-1.61$} & \multirow{6}{*}{$-0.40$} & \multirow{6}{*}{$-1.99$} \\
\textsc{bces} & $M_{\rm *,disk}$ & $-0.027\pm0.010$ & $10.69\pm0.04$ & 0.22 & 0.26 & & & & \\
\textsc{bces} & $|\phi|$ & $-0.081\pm0.043$ & $10.70\pm0.06$ & 0.30 & $0.38$ & & & & \\
\textsc{mpfitexy} & \textit{Symmetric} & $-0.066\pm0.018$ & $10.70\pm0.06$ & 0.26 & $0.33$ & & & \\
\textsc{mpfitexy} & $M_{\rm *,disk}$ & $-0.028\pm0.010$ & $10.70\pm0.04$ & $0.22$ & 0.26 & & & & \\
\textsc{mpfitexy} & $|\phi|$ & $-0.104\pm0.026$ & $10.69\pm0.08$ & $0.38$ & 0.47 & & & & \\
\hline
\multicolumn{11}{c}{21\tablenotemark{a} Early-type Galaxies with Core-S{\'e}rsic Bulges} \\
\hline
\multicolumn{11}{c}{The $M_{\rm BH}$--$M_{\rm *,tot}$ Relation: $\log(M_{\rm BH}/M_{\sun})=\alpha\log\left\{M_{\rm *,tot}/[\upsilon(2.58\times10^{11}\,M_{\sun})]\right\}+\beta$} \\
\hline
\textsc{bces} & \textit{Symmetric} & $1.34\pm0.19$ & $9.19\pm0.09$ & $0.37$ & $0.40$ & \multirow{6}{*}{$\left\}\begin{tabular}{@{}l@{}} \\ \\ \\ \\ \\ \\ \end{tabular}\right.$} & \multirow{6}{*}{$0.68$} & \multirow{6}{*}{$-3.12$} & \multirow{6}{*}{$0.63$} & \multirow{6}{*}{$-2.66$} \\
\textsc{bces} & $M_{\rm BH}$ & $0.96\pm0.22$ & $9.16\pm0.10$ & $0.34$ & 0.38 & & & & \\
\textsc{bces} & $M_{\rm *,tot}$ & $1.92\pm0.43$ & $9.25\pm0.10$ & 0.48 & $0.52$ & & & & \\
\textsc{mpfitexy} & \textit{Symmetric} & $1.32\pm0.23$ & $9.19\pm0.07$ & 0.37 & $0.40$ & & & \\
\textsc{mpfitexy} & $M_{\rm BH}$ & $0.95\pm0.25$ & $9.15\pm0.09$ & $0.34$ & 0.38 & & & & \\
\textsc{mpfitexy} & $M_{\rm *,tot}$ & $1.90\pm0.45$ & $9.24\pm0.12$ & $0.48$ & 0.52 & & & & \\
\enddata
\tablecomments{Late-type galaxies are from this work, and early-type galaxies are from \citet{Savorgnan:2016:II}. The calculation of the total rms scatter ($\Delta_{\rm rms}$), the correlation coefficients ($r$ and $r_s$), and their associated probabilities, do not take into account the uncertainties on the datapoints.
Column (1): regression software used.
Column (2): variable that had its offsets from the regression line minimized.
Column (3): slope.
Column (4): intercept.
Column (5): intrinsic scatter in the vertical $Y$-coordinate direction \citep[][their Equation~(1)]{Graham:Driver:2007}.
Column (6): total rms scatter in the $Y$-coordinate direction.
Column (7): Pearson correlation coefficient.
Column (8): logarithm of the Pearson correlation probability value.
Column (9): Spearman rank-order correlation coefficient.
Column (10): logarithm of the Spearman rank-order correlation probability value.
}
\tablenotetext{a}{This number was 22 in \citet{Savorgnan:2016:II} because they considered NGC~4594 to have a core-S{\'e}rsic bulge (and not to be a spiral galaxy).}
\end{deluxetable*}

\begin{figure}
\includegraphics[clip=true,trim= 0mm 0mm 0mm 0mm,width=\columnwidth]{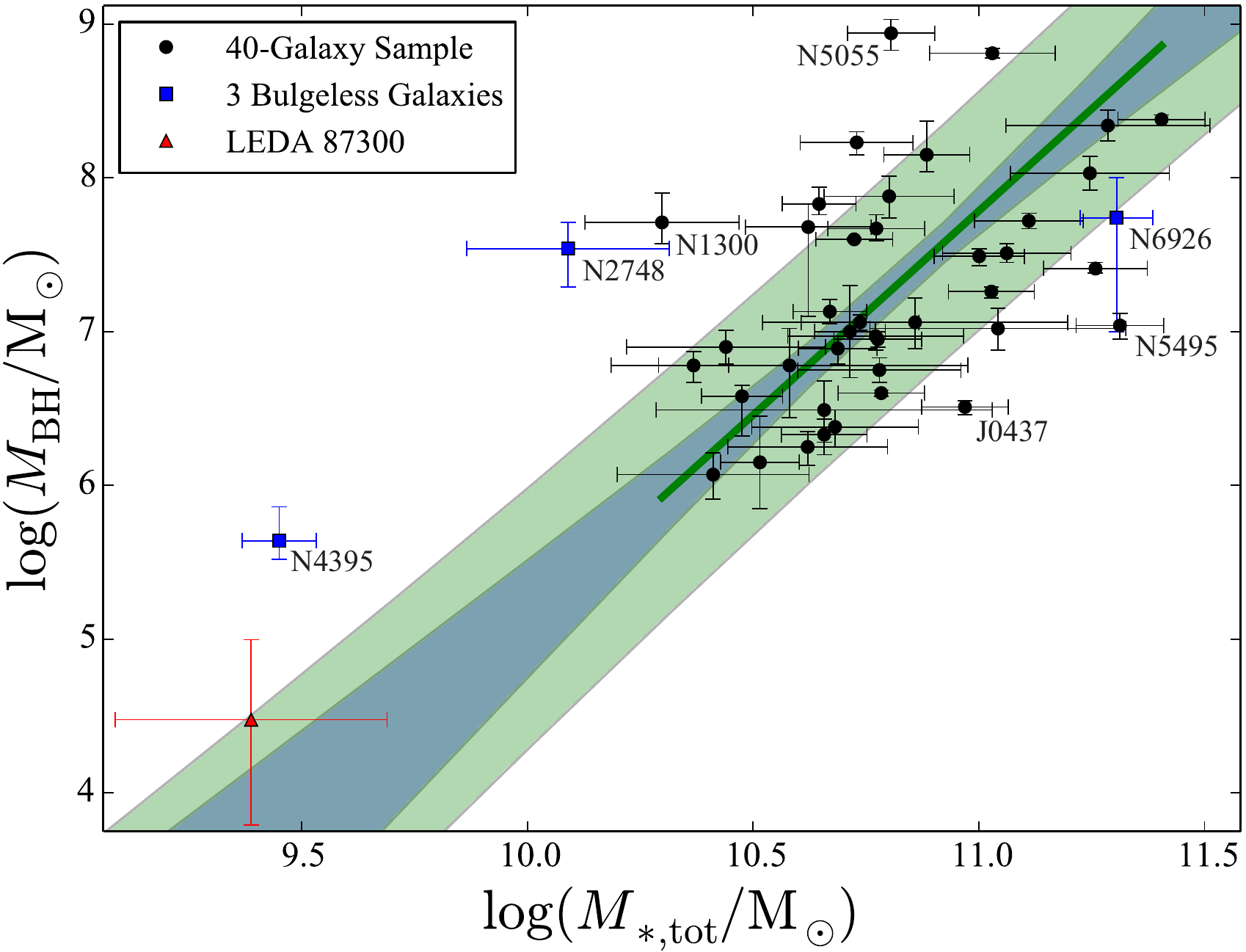}
\caption{Similar to Figure \ref{Ewan_M_BH-M_tot_plot}, except here we also
plot (but do not include in the regression) the three bulgeless galaxies from
our sample and the bulgeless galaxy LEDA~87300 \citep{Graham:2016}.  Here, we
plot the \textsc{mpfitexy} \textit{bisector} regression (solid green
line). The dark-green band shows the $\pm1$\,$\sigma$ uncertainty on the slope and
the intercept from the regression, while the light-green band delineates the
$\pm1$\,$\sigma$ scatter of the data about the regression line.}
\label{total_plot2}
\end{figure}

In Figure \ref{total_plot2}, we present the data slightly differently than in
Figure \ref{Ewan_M_BH-M_tot_plot}: we plot (but do not include in the
regression)\footnote{If the three bulgeless spiral galaxies are included in
the regression analysis, the \textsc{bces} \textit{bisector} routine finds a
slope of $2.11\pm0.37$. This slope is only $69\%$ as steep as the 40-galaxy slope;
its shallowness is strongly influenced by the position of NGC~4395. Such a
shallow slope is uncharacteristic, given that it is not steeper than the
$M_{\rm BH}$--$M_{\rm *,sph}$ relation.} the positions of the three excluded
bulgeless galaxies from our sample and the bulgeless galaxy LEDA~87300 \citep{Graham:2016}. Notably, our
extrapolated \textsc{mpfitexy} \textit{bisector} linear regression coincides
with the location of LEDA~87300, while NGC~4395 is an outlier. 

\subsubsection{The $M_{\rm BH}$--$M_{\rm *,disk}$ Relation}\label{sec:M_BH-M_disk}

\begin{figure}
\centering
\includegraphics[clip=true,trim= 4mm 4mm 1mm 1mm,width=\columnwidth]{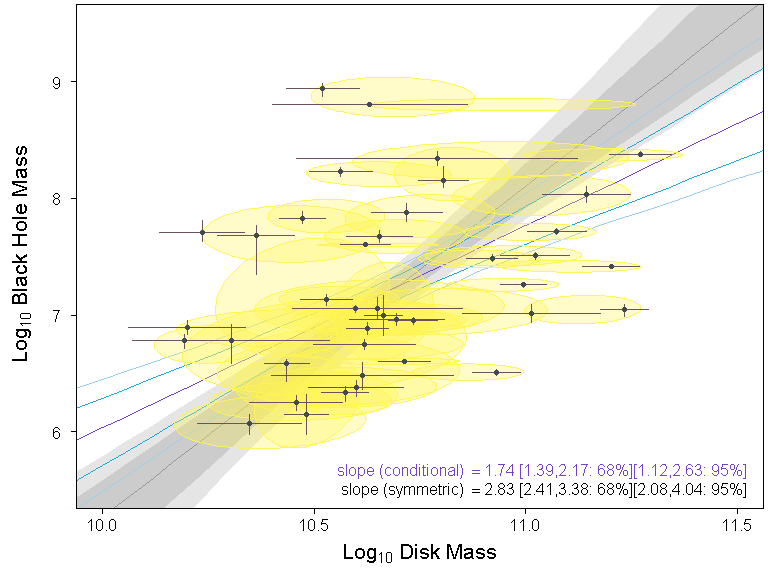}
\caption{Similar to Figure~\ref{Ewan_M_BH-M_tot_plot},
except that the disk stellar mass is plotted along the horizontal axis. The
gray line is represented by Equation~(\ref{M_BH-M_disk_eqn}). Masses are in units of solar masses.}

\label{Ewan_M_BH-M_disk_plot}
\end{figure}

The $\left(\log M_{\rm *,disk},\, \log M_{\rm BH}\right)$ data set has 
$r = 0.28$, $p=8.13\times10^{-2}$,
$r_s=0.34$, and $p_s=3.06\times10^{-2}$. However, as noted before, this does
not take into consideration the errors associated with the datapoints. Using
the \textit{symmetric} Bayesian analysis, we find
\begin{IEEEeqnarray}{rCl}
\log\left( \frac{M_{\rm BH}}{M_{\sun}}\right ) & = & \left(2.83_{-0.42}^{+0.55}\right)\log\left[ \frac{M_{\rm *,disk}}{\upsilon(4.98\times10^{10}\,M_{\sun})}\right ] \nonumber \\
&& +\> \left(7.24\pm0.13\right),
\label{M_BH-M_disk_eqn}
\end{IEEEeqnarray}
with $\Delta_{\rm rms} = 0.91$\,dex and $\epsilon=0.78$\,dex in the
$\log{M_{\rm BH}}$ direction (see
Figure~\ref{Ewan_M_BH-M_disk_plot}). The \textit{conditional} Bayesian
analysis, which minimizes the offsets of the (error-weighted) data in the
$\log{M_{\rm BH}}$ direction, has $\Delta_{\rm rms} = 0.75$\,dex
and $\epsilon=0.67$\,dex (see Table~\ref{table:bent}).

\begin{figure}
\includegraphics[clip=true,trim= 0mm 0mm 0mm 0mm,width=\columnwidth]{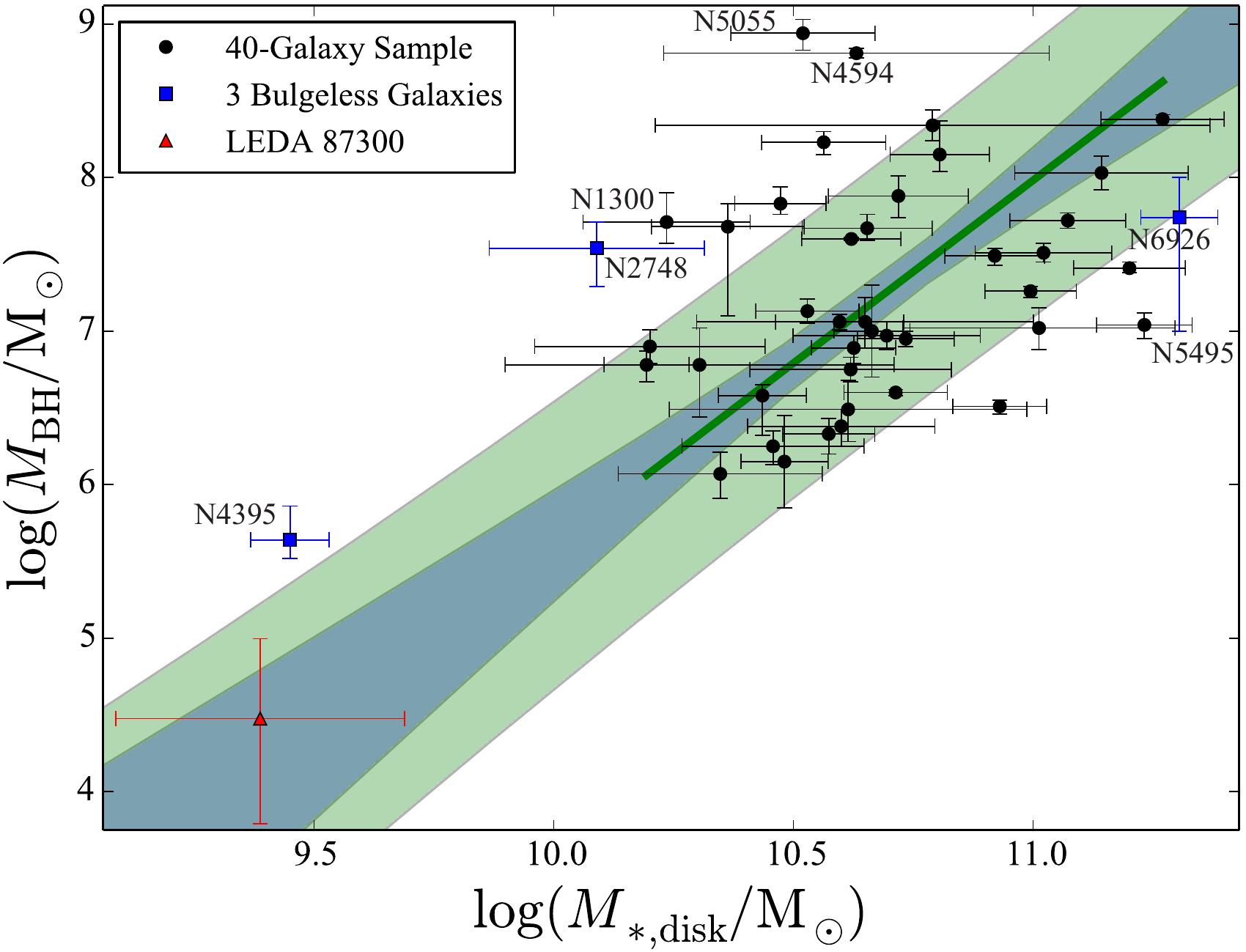}
\caption{Similar to Figure \ref{total_plot2}, except that the stellar disk
mass is plotted along the horizontal axis. Note that $M_{\rm *,disk}\equiv
M_{\rm *,tot}$ for the bulgeless galaxies (including LEDA~87300) that are
shown here, but they were excluded from the linear regression analysis (see Table~\ref{table:bent}).}
\label{disk_plot2}
\end{figure}

In Figure \ref{disk_plot2}, we plot (but do not include in the regression) 
the three bulgeless galaxies that were excluded from our sample, as well as the
bulgeless galaxy LEDA~87300 \citep[with masses taken from][]{Graham:2016}.
LEDA~87300 is consistent with the extrapolation of
our \textsc{mpfitexy} \textit{bisector} linear regression to lower masses,
while NGC~4395 is a slight outlier. 

\subsection{Relations with the Spiral Arm Pitch Angle ($\phi$)}

Nearly four decades ago, \citet{Kennicutt:1981} presented preliminary
evidence that spiral arm pitch angle is correlated with $M_{\rm
*,tot}$. Specifically, in his Figures~9 and 10, he illustrates a trend in both
the $\phi$--(absolute $B$-band galaxy magnitude) and the $\phi$--(maximum
rotational velocity) diagrams, respectively. With both of these quantities as
indicators of total galaxy mass, it is not unexpected that we should recover a
correlation between the pitch angle and the total stellar mass of a galaxy.

\begin{figure}
\centering
\includegraphics[clip=true,trim= 0mm 0mm 0mm 0mm,width=\columnwidth]{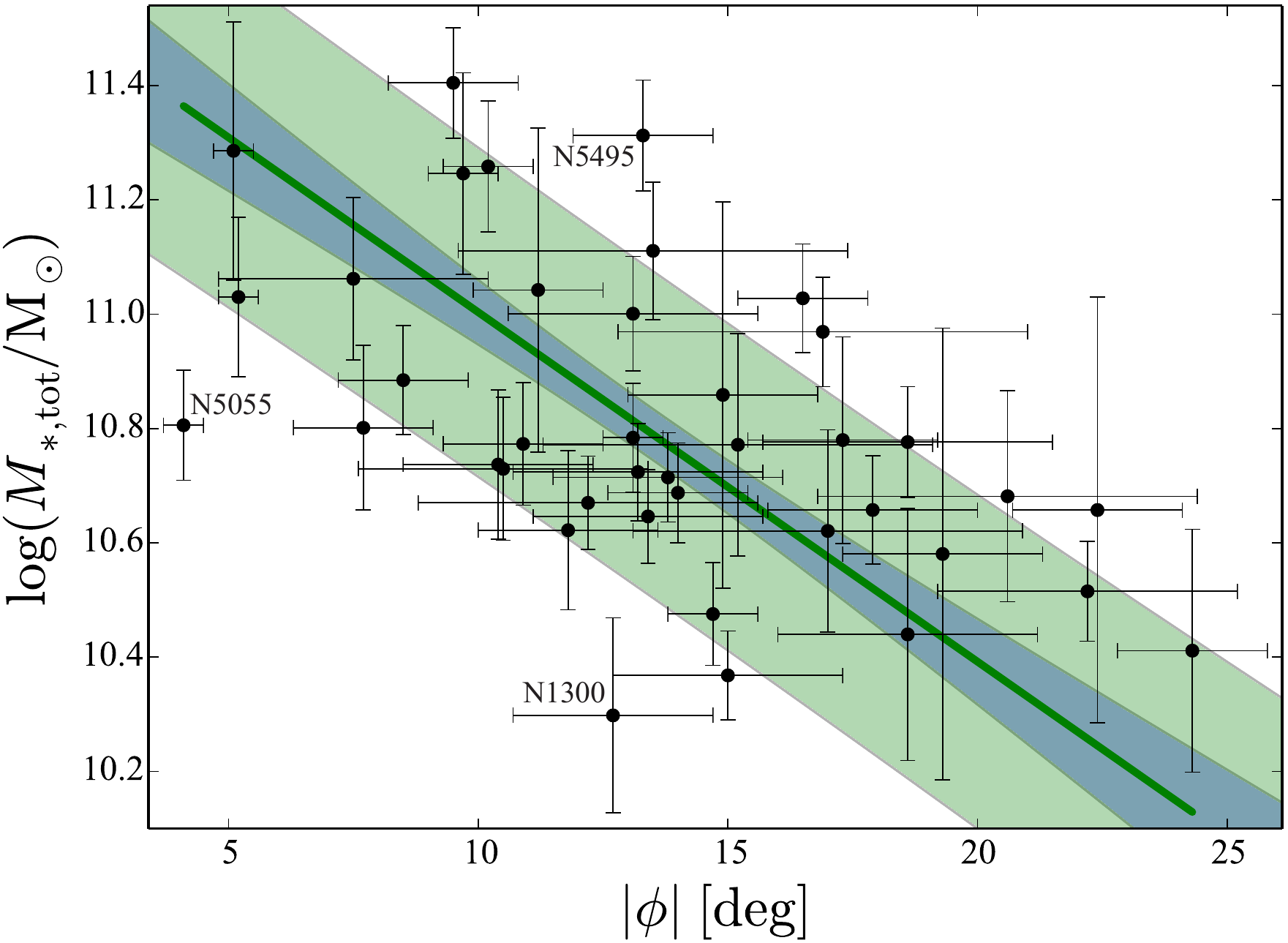}
\caption{Logarithmic spiral arm pitch angle versus the galaxy total stellar
mass. The \textsc{mpfitexy} \textit{bisector} regression is presented (see Table~\ref{table:bent}).}
\label{phi-M_tot_plot}
\end{figure}

\begin{figure}
\centering
\includegraphics[clip=true,trim= 0mm 0mm 0mm 0mm,width=\columnwidth]{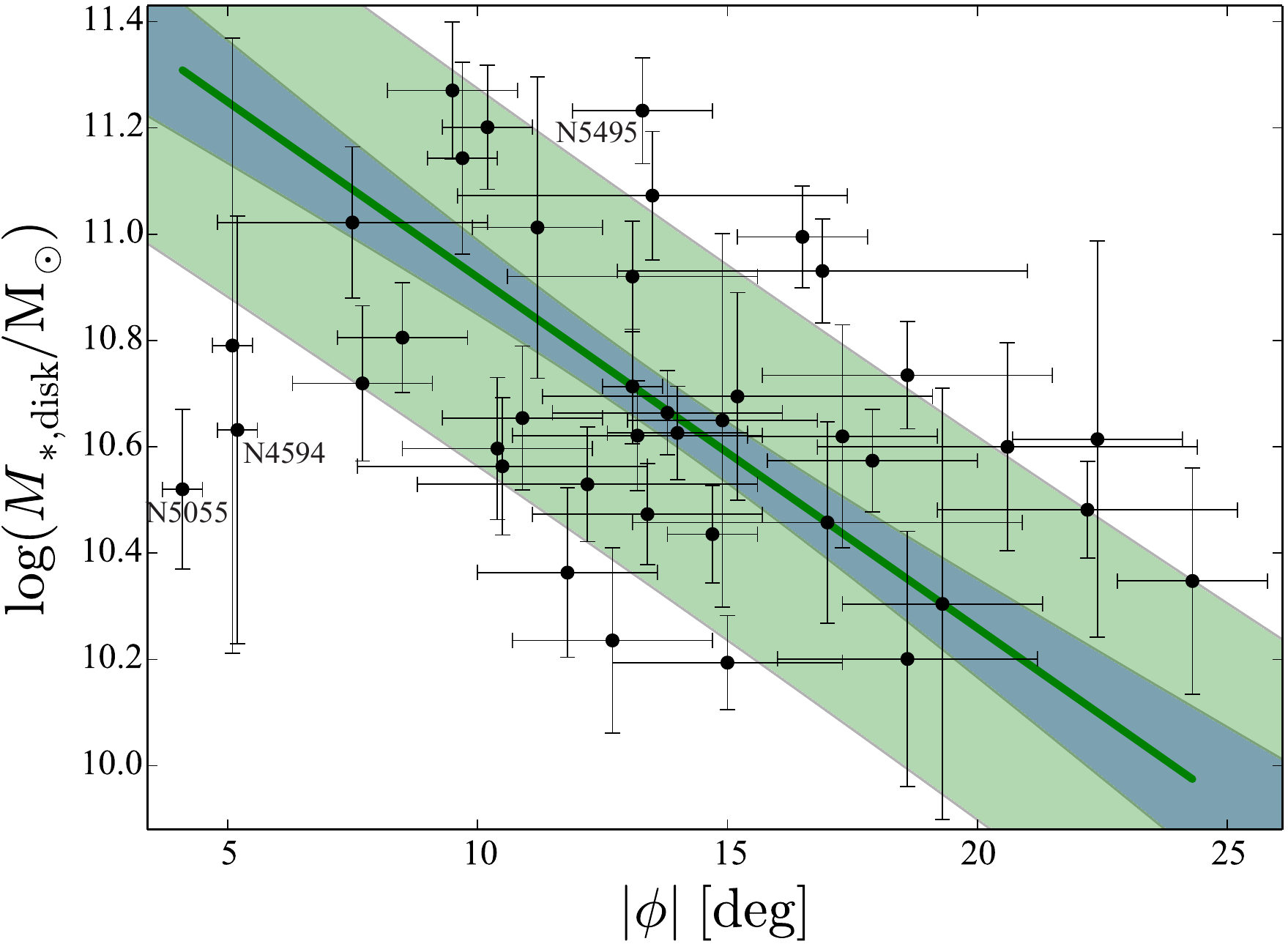}
\caption{Logarithmic spiral arm pitch angle versus the disk stellar
mass. The \textsc{mpfitexy} \textit{bisector} regression is presented (see Table~\ref{table:bent}).}
\label{phi-M_disk_plot}
\end{figure}

Since logarithmic spiral arm pitch angle ($\phi$) has been shown to correlate
well with black hole mass \citep{Seigar:2008,Berrier:2013,Davis:2017}, it is
prudent to check on the $M_{\rm *,tot}$--$\phi$ relation. We stress that the pitch angles are measured after first reprojecting the disks to a face-on orientation, and thus recovering the intrinsic geometry of the spiral arms. We additionally
explore the possibility of a relation existing between $M_{\rm
*,disk}$ and $\phi$, given that the spiral pattern resides in the disk, and the bulk of a
spiral galaxy's stellar mass is in its disk component. We present the diagrams
for the $M_{\rm *,tot}$--$\phi$ and $M_{\rm *,disk}$--$\phi$ relations in
Figures \ref{phi-M_tot_plot} and \ref{phi-M_disk_plot}, respectively, and the
results are presented in Table~\ref{table:bent}. As was the case with the
black hole mass relations, the stellar disk mass displays the weaker
correlation among these two comparisons with pitch angle.

\vspace{5mm}

\section{Discussion}\label{DI}

\subsection{From Bulge to Total Galaxy Mass}\label{sec:BT}

\citet{Lasker:2014} reported agreement between their $M_{\rm BH}$--$L_{\rm
*,sph}$ and $M_{\rm BH}$--$L_{\rm *,tot}$ relations, although their slopes are
much shallower (both less than 1) than our slopes for the $M_{\rm
BH}$--$M_{\rm *,sph}$ and $M_{\rm BH}$--$M_{\rm *,tot}$
relations (greater than 2 and 3, respectively) for spiral galaxies.  However, their sample of 35
galaxies contained only four spiral galaxies and, as such, cannot so readily be compared to
our analysis of 40 spiral galaxies.  
\citet{Lasker:2014} also reported 
consistent intrinsic scatter between their $M_{\rm BH}$--$L_{\rm *,sph}$ and
$M_{\rm BH}$--$L_{\rm *,tot}$ relations, whereas \citet{Savorgnan:2016:II}
found from their sample of 66 galaxies (including 17 spiral galaxies) that the
claim of \citet{Lasker:2014} is only valid for (bright) early-type galaxies.
In N.\ Sahu et al.\ (2019, submitted), we will provide the results from our analysis
of $\approx80$ early-type galaxies with directly measured black hole masses, 
building on \citet{Lasker:2014} and \citet{Savorgnan:2016:II}. 

As for spiral galaxies with S\'ersic bulges, when comparing the estimated
intrinsic scatters from our various linear regressions, we find that the
median intrinsic scatter for the $M_{\rm BH}$--$M_{\rm *,sph}$ relation
is 0.18\,dex \emph{less} than that of the $M_{\rm
BH}$--$M_{\rm *,tot}$ relation. Contrary to this, for a sample of 21
early-type galaxies with core-S{\'e}rsic bulges, taken
from \citet{Savorgnan:2016:II}, we find that the median intrinsic scatter for
the $M_{\rm BH}$--$M_{\rm *,sph}$ relation is 0.05\,dex \emph{more} than that of the $M_{\rm BH}$--$M_{\rm *,tot}$ relation.
However, it should be borne in mind that the slope increases notably when
going from the $M_{\rm BH}$--$M_{\rm *,sph}$ to the $M_{\rm BH}$--$M_{\rm
*,tot}$ relation for late-type galaxies and roughly stays the same for
early-type galaxies with core-S{\'e}rsic bulges. The increase of slope
naturally causes the scatter to also increase in the vertical direction, i.e.,
along the black hole mass axis.  This complicates the simple comparison of
intrinsic scatter across scaling relations with various slopes.

We find a correlation between black hole mass and the total stellar mass of
spiral galaxies that is not as strong ($r=0.47$ and $r_s=0.53$) as the correlation
between black hole mass and bulge stellar mass ($r=0.66$ and $r_s=0.62$). The
rms scatter in the $\log{M_{\rm BH}}$ direction from
the \textit{conditional} Bayesian linear regression, about the $M_{\rm
BH}$--$M_{\rm *,tot}$ relation, is 0.66\,dex (cf.\ 0.60\,dex
for the $M_{\rm BH}$--$M_{\rm *,sph}$ relation). The \textit{symmetric}
Bayesian analysis slope $(3.05_{-0.49}^{+0.57})$ is consistent
with the \textsc{bces} $(3.05\pm0.70)$ and \textsc{mpfitexy}
$(2.65\pm0.65)$ \textit{bisector} slopes at the level of $0.00$\,$\sigma$ and $0.35$\,$\sigma$, respectively. Likewise,
the \textit{conditional} Bayesian analysis
slope $(2.03_{-0.41}^{+0.44})$ is consistent with
the \textsc{bces} $(2.04\pm0.73)$ and \textsc{mpfitexy} $(1.62\pm0.39)$
$(Y|X)$ slopes at the level of 0.01\,$\sigma$ and $0.51$\,$\sigma$, respectively.

\begin{figure}
\includegraphics[clip=true,trim= 0mm 0mm 0mm 0mm,width=\columnwidth]{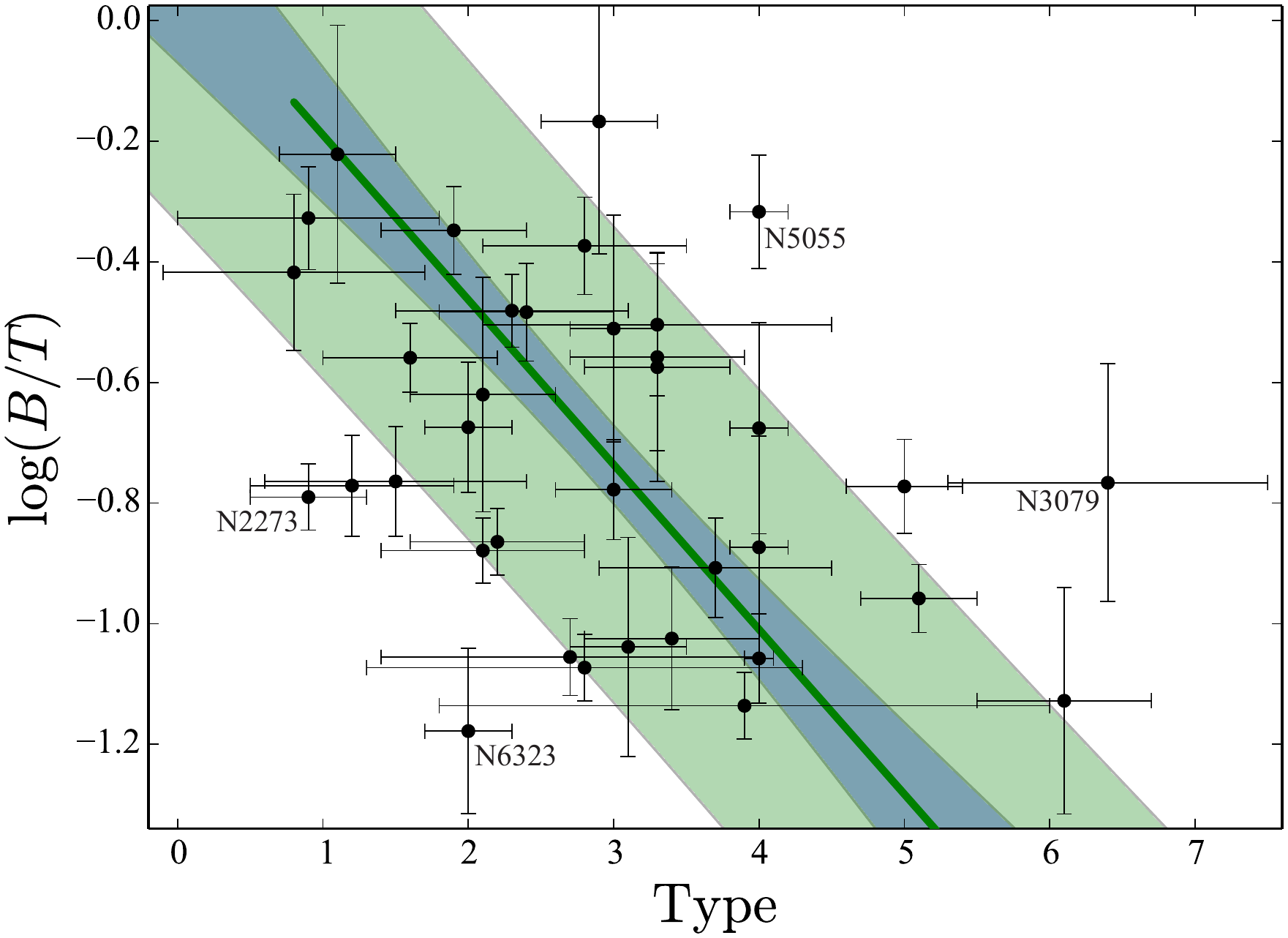}
\caption{Logarithm of the bulge-to-total flux ratio versus the numerical
morphological type (for 36 spiral galaxies from our sample with both measurements), with Equation (\ref{T-log_B_T_eqn}) plotted.}
\label{fig:T-log_B_T}
\end{figure}

Even though statistically equivalent (at the level of $0.73$\,$\sigma$), the slope of our $M_{\rm BH}$--$M_{\rm *,tot}$ relation
(Equation \ref{Ewan_M_BH-M_tot_eqn}) is noticeably
($25\%$) steeper than that of our $M_{\rm BH}$--$M_{\rm
*,sph}$ relation \citepalias[][equation~12]{Davis:2018}. Because the bulge-to-total $(B/T)$ flux ratio changes with the morphological type of spiral galaxies, as do
the black hole masses, one does not expect $M_{\rm BH}$ vs.\ $T$ to have the same
slope as $M_{\rm BH}$ vs.\ $B$. In Figure~\ref{fig:T-log_B_T}, we explore this
by first demonstrating that there indeed is a trend between the $B/T$ flux
ratio and the numerical morphological type; earlier types with more massive bulges have
greater $B/T$ ratios,\footnote{This is consistent with the quantitative studies
of \citet{Graham:Worley:2008} and largely driven by the changing bulge flux
with spiral galaxy type 
\citep[][, their Figures~1 and 2]{Yoshizawa_Wakamatsu:1975}.} such that
\begin{equation}
\log\left(\frac{B}{T}\right ) = -(0.27\pm0.08)[\rm{Type}-2.85] - (0.70\pm0.06),
\label{T-log_B_T_eqn}
\end{equation}
with $\Delta_{\rm rms} = 0.37$\,dex and $\epsilon=0.31$\,dex in the $\log(B/T)$ direction from the \textsc{bces} \textit{bisector} regression; $r = -0.37$, $p$ = $2.73\times10^{-2}$, $r_s=-0.35$, and $p_s$ = $3.71\times10^{-2}$.

\begin{figure}
\includegraphics[clip=true,trim= 0mm 0mm 0mm 0mm,width=\columnwidth]{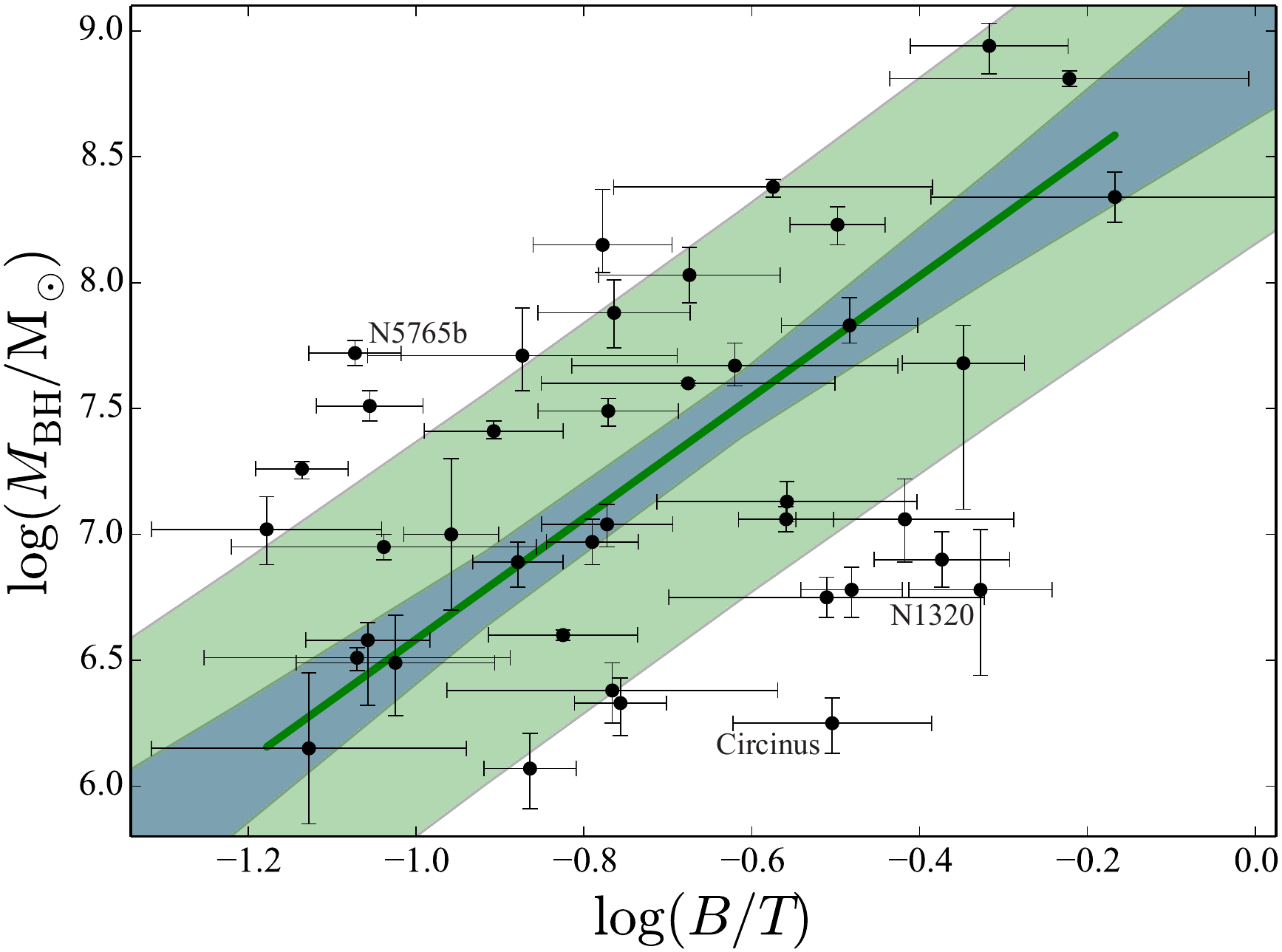}
\caption{SMBH mass vs.\ the difference between the bulge and total flux, with Equation (\ref{M_BH-Delta_M_eqn}) plotted.}
\label{total_study}
\end{figure}

We additionally reveal how the $B/T$ flux ratio changes with the black hole
mass. In Figure~\ref{total_study}, we show that the largest SMBHs (which
typically reside in the largest bulges) have the largest $\log(B/T)$ values,
thus confirming that the $M_{\rm BH}$--$M_{\rm *,tot}$ relation should be
steeper than the $M_{\rm BH}$--$M_{\rm *,sph}$ relation.  We find from the \textsc{bces} \textit{bisector} analysis that
\begin{IEEEeqnarray}{rCl}
\log\left( \frac{M_{\rm BH}}{M_{\sun}}\right ) & = & (2.41\pm0.46)\log\left[ \frac{\log(B/T)}{-0.77}\right ] \nonumber \\
&& +\> (7.15\pm0.12),
\label{M_BH-Delta_M_eqn}
\end{IEEEeqnarray} 
with $\Delta_{\rm rms} = 0.73$\,dex and $\epsilon=0.69$\,dex in the $\log{M_{\rm
BH}}$ direction; $r = 0.43$, $p$ = $5.43\times10^{-3}$, $r_s=0.35$, and $p_s$ = $2.58\times10^{-2}$.

\begin{figure}
\includegraphics[clip=true,trim= 8mm 7mm 20mm 15mm,width=\columnwidth]{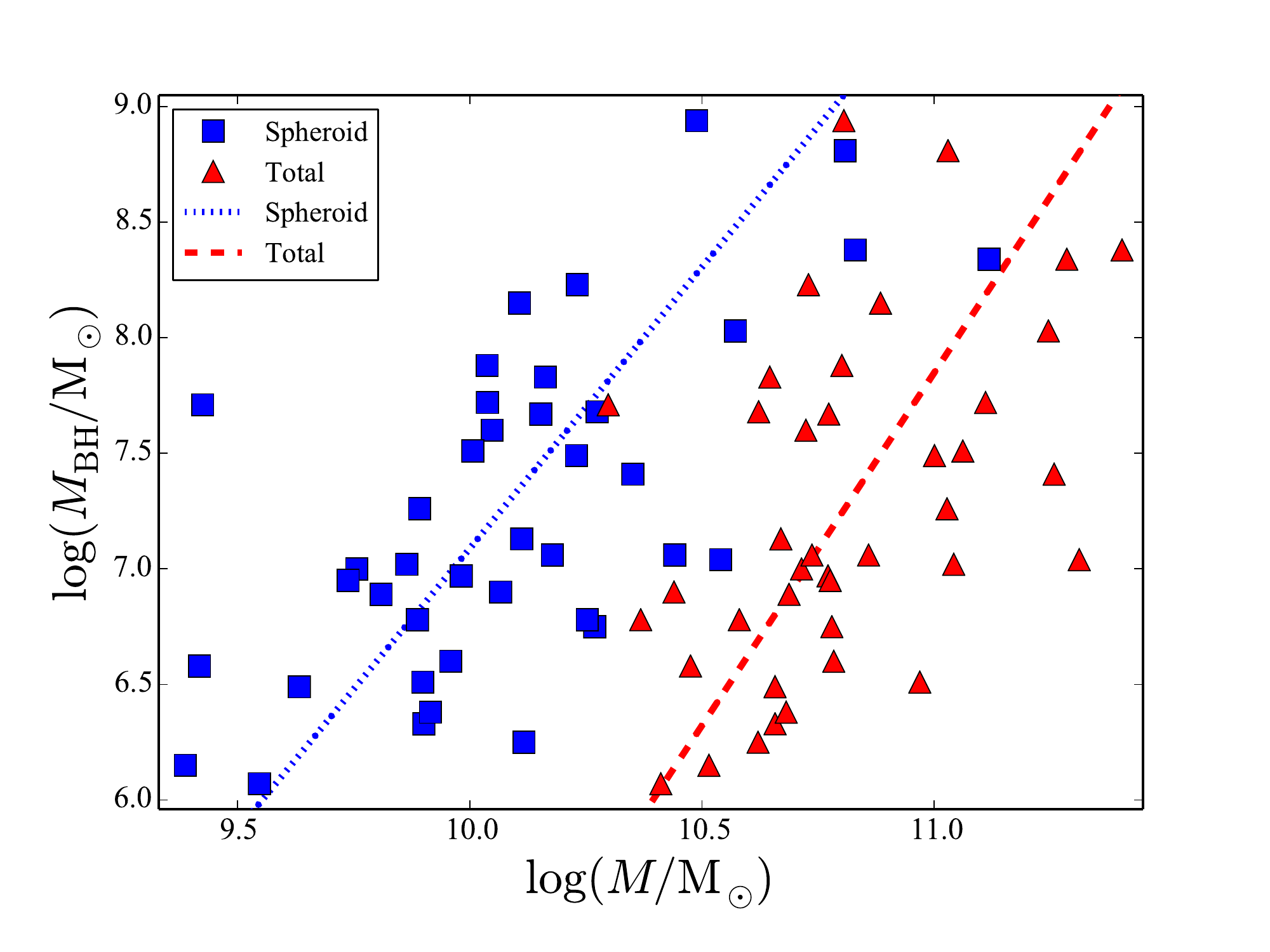}
\caption{This figure combines the data from Figure~5 in \citetalias{Davis:2018} with Figure~\ref{Ewan_M_BH-M_tot_plot} from this work, represented with blue and red, respectively. Shifting the total galaxy stellar masses (\textcolor{red}{red} triangles) left by an amount equal to $\log(B/T)$, transforms them into the spheroid stellar masses (\textcolor{blue}{blue} squares). The dotted \textcolor{blue}{blue} line and the dashed \textcolor{red}{red} line represent the \textit{symmetric} Bayesian regression lines (equation~12 from \citetalias{Davis:2018} with Equation~\ref{Ewan_M_BH-M_tot_eqn} from this work) for the spheroid and total stellar masses, respectively. Note that errors on individual points and on the fitted lines have been omitted for clarity.}
\label{M_BH-M_plot}
\end{figure}

In Figure~\ref{M_BH-M_plot}, we demonstrate that the $M_{\rm BH}$--$M_{\rm
*,tot}$ relation has a steeper slope than the $M_{\rm BH}$--$M_{\rm *,sph}$
relation in \citetalias{Davis:2018}, which can be understood via the morphological relations
given above. Similarly, the $M_{\rm *,tot}$--$\phi$ relation (Figure~\ref{phi-M_tot_plot} and Table~\ref{table:bent}) 
possesses a shallower slope than the $M_{\rm *,sph}$--$\phi$ relation in \citetalias{Davis:2018}. The $M_{\rm *,tot}$--$\phi$ relation's shallowness is opposite to the $M_{\rm BH}$--$M_{\rm *,tot}$ relation's steepness because pitch angle is
anticorrelated with black hole mass.\footnote{In the absence of uncertainty on $M_{\rm BH}$ or $\phi$, the slopes for the various relations will be such that $M_{\rm BH}$--$M_{\rm *,sph}$ $<$ $M_{\rm BH}$--$M_{\rm *,tot}$ $<$ $M_{\rm BH}$--$M_{\rm *,disk}$ and $M_{\rm *,sph}$--$\phi$ $>$ $M_{\rm *,tot}$--$\phi$ $>$ $M_{\rm *,disk}$--$\phi$. This can be seen by comparing the various \textit{conditional} regressions that minimize the offsets with $M_{\rm *,sph}$, $M_{\rm *,tot}$, or $M_{\rm *,disk}$ from \citetalias{Davis:2018} and this work.}
In passing, we note that we did explore the expected trend between black hole mass and galaxy color, but
the overwhelming majority of spiral galaxies with directly measured black hole
masses have red $B-K$ colors, prohibiting the usefulness of this particular
diagram at this stage.

Since our galaxies are disk dominated, the strong $M_{\rm BH}$--$M_{\rm
*,sph}$ relation and weak $M_{\rm BH}$--$M_{\rm *,disk}$ relation suggest
that the $M_{\rm BH}$--$M_{\rm *,tot}$ relation is governed mainly by the
influence of the $M_{\rm BH}$--$M_{\rm *,sph}$ relation. 
While the latter relation may be more fundamental, 
the correlation between black hole mass and total galaxy stellar mass is
probably more useful. It provides an easy and quick way to estimate central black hole mass
in spiral galaxies by simply measuring the total luminosity and then
converting into stellar mass.

Our presentation of the
$M_{\rm BH}$--$M_{\rm *,disk}$ relation is primarily to demonstrate that black
holes are not unrelated to properties of their galactic disks, which is partly 
reinforced by a strong correlation with the winding geometry of the spiral arms
(which live in the disk). 
For late-type spiral galaxies, which have low bulge-to-disk ($B/D$) flux
ratios compared to early-type spiral galaxies, the disk constitutes the
majority of the total galaxy mass (see Figure \ref{fig:T-log_B_T}). This
implies that if the SMBH mass correlates with the total stellar mass --- which
need not be a direct correlation --- then it 
should also correlate with the disk stellar mass. However, one can also appreciate
how sample selection can result in one not finding this correlation: a small
range of disk stellar masses, or a small number of galaxies, or poor disk
magnitudes from the galaxy decomposition will hinder success.

\subsection{Potential Over/undermassive Black Holes}

Figures \ref{total_plot2} and \ref{disk_plot2} reveal that NGC~1300 and
NGC~5055 are outliers above the $M_{\rm BH}$--$M_{\rm *,tot}$ and $M_{\rm
BH}$--$M_{\rm *,disk}$ lines.  Either their total/disk masses are lower than
expected or their black hole masses are higher than expected.  While NGC~5055
(also known as M63 or the ``Sunflower Galaxy'') appears to have a slightly
overmassive black hole in the $M_{\rm BH}$--$M_{\rm *,tot}$ and $M_{\rm
BH}$--$M_{\rm *,disk}$ diagrams, it does not in the $M_{\rm BH}$--$M_{\rm
*,sph}$ diagram in \citetalias{Davis:2018}.  However, \citet{Davis:2017} revealed that NGC~5055 is a
prominent outlier in the $M_{\rm BH}$--$\sigma_*$ diagram (where $\sigma_*$ is
the stellar velocity dispersion), indicating a possible overmassive black
hole in this galaxy.  NGC~1300 stands out as a quintessential example of a
strongly barred spiral galaxy with nuclear spiral arms; it is the least massive galaxy in our sample, yet its black hole appears to be overmassive by $\approx1.5$\,dex. Finally, NGC~5495 is
an outlier in most of the diagrams. Of our 40-galaxy sample, it has the second-highest $M_{\rm *,tot}$ and $M_{\rm *,disk}$. However, its black hole seems to
be undermassive by $\approx1.5$\,dex. NGC~1300 and NGC~5495 are outliers in all three relations: $M_{\rm BH}$--$M_{\rm *,sph}$, $M_{\rm BH}$--$M_{\rm *,tot}$, and $M_{\rm BH}$--$M_{\rm *,disk}$.

\subsection{Relations with the Spiral Arm Pitch Angle ($\phi$)}

As with the $M_{\rm BH}$--$M_{\rm *,tot}$ relation
(Figures \ref{Ewan_M_BH-M_tot_plot} and \ref{total_plot2}), the $M_{\rm
*,tot}$--$\phi$ relation (Figure \ref{phi-M_tot_plot}) also displays a similarly
correlated fit. Since our galaxies are mainly disk dominated (their median
bulge-to-total flux ratio is $0.17$), this implies that at least two
properties of the disk (its stellar mass and pitch angle) should be correlated
with the black hole mass. Furthermore, since the pitch angle correlates well
with the SMBH mass \citep{Seigar:2008,Berrier:2013,Davis:2017} plus bulge mass
and total mass (figure~8 from \citetalias{Davis:2018} and Figure~\ref{phi-M_tot_plot} from
this work), there should be a correlation between $M_{\rm *,disk}$ and $\phi$,
as demonstrated in Figure \ref{phi-M_disk_plot}.

The strength of the correlation between $M_{\rm *,disk}$ and $\phi$ is less than that between $M_{\rm *,sph}$ and $\phi$; the Pearson correlation coefficients are $-0.35$ and $-0.63$, respectively. This may seem unexpected, as the spiral arms are a feature of the disk. However, it should be remembered that the spiral density wave depends on the density of the disk, rather than the total mass of the disk, and it is the mass of the bulge that effectively anchors the spiral arm, a bit like setting the tension in the vibrating string of a violin by adjusting the tuning peg \citep{Davis:2015}.

\subsection{Morphology-dependent $M_{\rm BH}$--$M_{\rm *,tot}$ Relations}\label{sec:bent}

\begin{figure}
\includegraphics[clip=true,trim= 10mm 6mm 20mm 15mm,width=\columnwidth]{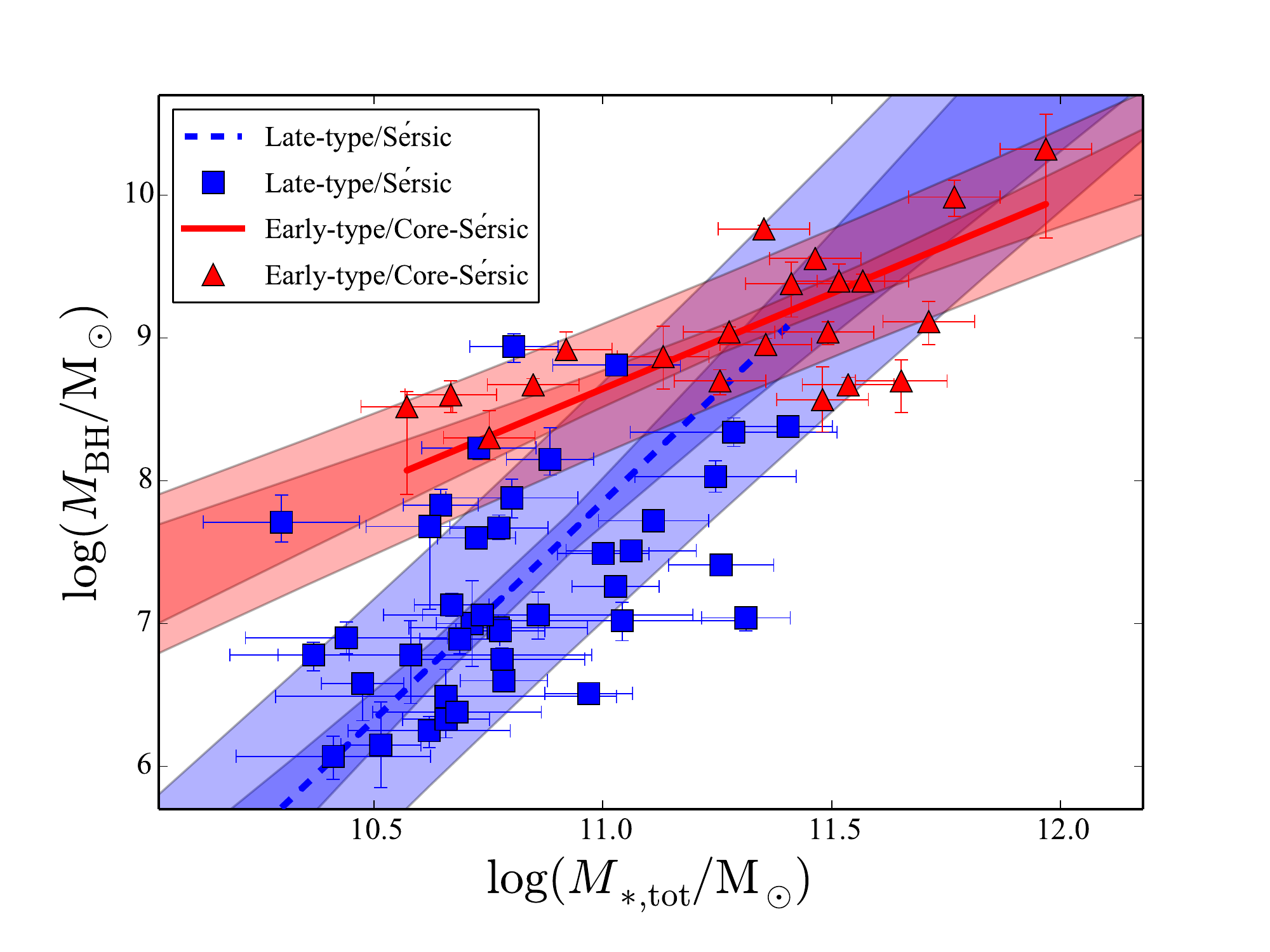}
\caption{Comparison of $M_{\rm BH}$ vs.\ $M_{\rm *,tot}$ for our 40 late-type/S{\'e}rsic galaxies and 21 early-type/core-S{\'e}rsic galaxies from \citet{Savorgnan:2016:II}. Note that all trend lines are from the \textsc{bces} \textit{bisector} routine.}
\label{bent_plots}
\end{figure}

For comparison, we show (in Figure~\ref{bent_plots}) how the $M_{\rm BH}$--$M_{\rm *,tot}$ relation appears when
generated from a sample of early-type galaxies with core-S{\'e}rsic spheroids
(which have black hole masses greater than $10^8\,M_{\sun}$) --- thought to have been built from major dry merger events. We obtained
measurements for a sample of 21 such galaxies
from \citet{Savorgnan:2016:II}. By analyzing that sample separately from
ours, we show that the slope for early-type core-S{\'e}rsic
galaxies in the $M_{\rm BH}$--$M_{\rm *,tot}$ diagram  ($\approx1.33$) is half as steep as
the slope of the $M_{\rm BH}$--$M_{\rm *,tot}$ relation for our 40 spiral
galaxies (see Table~\ref{table:bent}).

Figure \ref{bent_plots} shows a dichotomy between the slopes of early-type core-S{\'e}rsic galaxies and late-type S{\'e}rsic galaxies. If we compare the \textsc{bces} \textit{bisector} slopes of the $M_{\rm BH}$--$M_{\rm *,tot}$ relation for the early-type ($1.34\pm0.19$) and late-type ($3.05\pm0.70$) galaxies, we find that they are statistically different, agreeing only at the level of $1.92$\,$\sigma$. This illustrates that the two samples are fundamentally different. Fitting a single power law to the combined sample yields a slope for the $M_{\rm BH}$--$M_{\rm *,tot}$ relation of $2.90\pm0.21$ (according to the \textsc{bces} \textit{bisector} routine). This is notably different from the slope of $1.71\pm0.10$ found in \citetalias{Davis:2018} from fitting a single linear regression to the combined sample of 61 galaxies for the $M_{\rm BH}$--$M_{\rm *,sph}$ relation. 

This clear difference in the relations between different morphological types
echoes the results found in \citet{Savorgnan:2016:II} and \citetalias{Davis:2018} concerning
the $M_{\rm BH}$--$M_{\rm *,sph}$ relation.  In addition to these physical
differences between samples of varying morphological types, important empirical ramifications exist for the study of black hole mass scaling
relations. Therefore, we advise
caution for studies of scaling relations concerning the demographics of one's
chosen sample.
This is not only true for local samples, where one needs to use the
appropriate relation when predicting black hole masses, but care must also be
given to evolutionary studies.  For example, if one compared the $M_{\rm
BH}$--$M_{\rm *,tot}$ relation from a local hybrid sample (of late- and
early-type galaxies) with that from a higher-redshift galaxy sample of
early-type galaxies, the scaling relations may differ solely as a result of the use of
different morphological types at different epochs. 

\subsection{Predicting Black Hole Masses}

Considering black hole mass scaling relations with $\phi$, $M_{\rm *,sph}$,
$M_{\rm *,tot}$, or $\sigma_{*}$, we advocate that $\phi$ be preferentially
utilized for spiral galaxies with clear spiral structure.  We say this based
on the small total rms scatter, of just 0.43\,dex in the $\log M_{\rm BH}$ direction, about the
shallow $M_{\rm BH}$--$\phi$ relation \citep{Davis:2017}. 
For spiral galaxies 
without clear spiral structure, $M_{\rm *,sph}$ should be utilized, depending
on the desired accuracy and/or time requirements.  For bulgeless spiral
galaxies without clear spiral structure, $M_{\rm *,tot}$ can be
used. Importantly, use of $M_{\rm *,tot}$ has the clear advantage that it can be
measured for any spiral galaxy.  In passing, we also note that the measurement
of the stellar velocity dispersion $\sigma_{*}$ requires telescope-time-expensive
spectral data, while $M_{\rm *,sph}$ and $M_{\rm *,tot}$ just require photometric
data, but $\phi$ needs only a photometrically uncalibrated image.

The rms scatter in the $\log M_{\rm BH}$ direction is 0.60\,dex about
the $M_{\rm BH}$--$M_{\rm *,sph}$ relation and 0.66\,dex about
the $M_{\rm BH}$--$M_{\rm *,tot}$ relation, each from the \emph{conditional} Bayesian regressions. However, this quantity is not the
``be all and end all'' in deciding what relation is the most fundamental.  It
should be recognized that we have followed tradition and not advocated an error-weighted rms scatter,
and as such, outlying datapoints with small measurement errors will inflate this
reported scatter.  

Finally, our newly defined relations allow us to estimate which galaxies might
potentially harbor IMBHs ($10^2 \leq M_{\rm BH}/{\rm
M_{\sun}} \leq 10^5$). The \textit{symmetric} Bayesian analyses\footnote{It
would be a mistake to extrapolate the \textit{conditional} Bayesian line to
masses below the mass range used to construct it, because its shallow slope would
overestimate the black hole masses in this regime.} predict that 
galaxies with $M_{\rm *,tot}\leq\upsilon(1.16\times10^{10}\,M_{\sun})$
and/or $M_{\rm *,disk}\leq\upsilon(8.05\times10^9\,M_{\sun})$ should possess IMBHs.

In future work, we intend to explore the inclusion of additional parameters, 
which may potentially yield a tighter relation in the form of a 2D plane in a three-parameter space
rather than a 1D line in a two-parameter space. The increased spatial resolution\footnote{Enables smaller
spheres of influence to be measured.} and
sensitivity\footnote{Provides less noisy spectra and therefore better velocity
dispersions.} from the next generation of 20--30\,m class telescopes
will undoubtedly yield exciting results as one is afforded the ability
to probe a little deeper into the spiral galaxy (blue) sequence.
Already, advancements with interferometry like the Atacama Large
Millimeter/submillimeter Array (ALMA) are allowing one to achieve
angular resolutions as small as $0\farcs02$ (at 230\,GHz with the
16\,km baseline configuration).

An alternative avenue that we are
currently pursuing is the use of X-ray emission to detect the presence
of IMBHs in blue, late-type spiral galaxies (R.\ Soria et al.\ 2018, in
preparation). Over 50 spiral galaxies in the Virgo Cluster have recently
been observed with the Advanced CCD Imaging Spectrometer (ACIS-S)
detector, as a part of the 559\,ks {\it Chandra} Large Project titled ``Spiral
Galaxies of the Virgo Cluster'' (PI: R.~Soria; proposal ID: 18620568).
We will use the $M_{\rm BH}$--$M_{\rm *,tot}$ relation 
from this paper, as well as the $M_{\rm BH}$--$\phi$ relation from
\citet{Davis:2017}, to independently predict the black hole masses in these galaxies
\citep{Graham:2018}.

\section{Conclusions}\label{END}

This work built on many recent studies of black hole mass scaling relations
and has tried to advance the field by focusing on spiral galaxies with
detailed bulge, disk, etc., decompositions. This has allowed us to better
investigate the nature of the low-mass end of the black hole mass scaling
relations with unparalleled accuracy and greatly narrow down the uncertainty
on the slope of the $M_{\rm BH}$--$M_{\rm *,tot}$ relation for spiral
galaxies. We find the following significant results:

\begin{enumerate}
\item As expected, the $M_{\rm BH}$--$M_{\rm *,tot}$ slope is steeper than the
$M_{\rm BH}$--$M_{\rm *,sph}$ relation. We find $\log{M_{\rm
BH}}\propto\left(3.05_{-0.49}^{+0.57}\right )\log{M_{\rm *,tot}}$, while 
\citetalias{Davis:2018} found $\log{M_{\rm BH}}$ $\propto\left(2.44_{-0.31}^{+0.35}\right )$ $\log{M_{\rm
*,sph}}$ for the same sample of 40 spiral galaxies.
\item For large surveys, where accurate bulge/disk decompositions may not be
feasible, one may prefer to use the $M_{\rm BH}$--$M_{\rm *,tot}$ relation,
with its slightly greater rms scatter of $0.79$\,dex
(cf.\ 0.70\,dex about the $M_{\rm BH}$--$M_{\rm *,sph}$ relation) in the
$\log M_{\rm BH}$ direction when using the \textit{symmetric} regression. The scatter reduces to 0.66\,dex and 0.60\,dex, respectively, when using the asymmetric ($\textit{conditional}$) regression, which minimizes the scatter in only the $\log M_{\rm BH}$ direction.
\item It is advisable to not mix samples of early- and late-type galaxies. The slope of the $M_{\rm BH}$--$M_{\rm *,tot}$ relation for late-type galaxies is approximately twice as steep as that ($\approx1.3$) for early-type galaxies with core-S{\'e}rsic spheroids.
\item There \emph{is} a relation between black hole mass and disk mass. Although the
Spearman rank-order correlation coefficient is low, with $r_s=0.34$ and
$p_s=3.06\times10^{-2}$, this does note take into account the uncertainties on
the datapoints. Our \textit{symmetric} Bayesian analysis reveals a well-defined relation
(Equation~\ref{M_BH-M_disk_eqn}) with an $\approx 17\%$ uncertainty on the
slope. Furthermore, the low-mass bulgeless galaxy LEDA~87300 appears
consistent with this relation at $M_{\rm BH}=3.0\times10^4\,M_{\sun}$.
\item In Figures \ref{phi-M_tot_plot} and \ref{phi-M_disk_plot}, we provide the
relations between the spiral arm pitch angle ($\phi$) and the stellar mass of
the galaxy and disk (by which we include everything other than the
bulge). Given the strong correlation between $M_{\rm
BH}$ and $\phi$ \citep[e.g.,][]{Davis:2017}, these two relations draw strong parallels
with the two black hole mass scaling relations above.  That is, we have
checked and found consistency among these scaling relations. 
\end{enumerate}

Black hole mass scaling relations allow astronomers to quickly estimate black hole masses for large samples in an era of astrophysics research that is dominated by massive amounts of data. We present a refined $M_{\rm BH}$--$M_{\rm *,tot}$ relation for spiral galaxies, which is capable of producing expeditious, yet accurate, SMBH mass predictions.

\acknowledgments

We thank Nandini Sahu for her helpful comments and insights, which helped improve this paper. A.W.G. was supported under the Australian Research Council's funding scheme DP17012923. Parts of this research were conducted by the Australian Research Council Centre of Excellence for Gravitational Wave Discovery (OzGrav), through project no. CE170100004. This research has made use of NASA's Astrophysics Data System. This research has made use of the NASA/IPAC Infrared Science Archive. We acknowledge the usage of the HyperLeda database \citep{HyperLeda}, \url{http://leda.univ-lyon1.fr}. This research has made use of the NASA/IPAC Extragalactic Database (NED). Some of the data presented in this paper were obtained from the Mikulski Archive for Space Telescopes (MAST). This publication makes use of data products from the Two Micron All Sky Survey, which is a joint project of the University of Massachusetts and the Infrared Processing and Analysis Center/California Institute of Technology. The \textsc{bces} routine \citep{BCES} was run via the \textsc{python} module written by Rodrigo Nemmen \citep{Nemmen:2012}, which is available at \url{https://github.com/rsnemmen/BCES}. Error propagation calculations were performed via the \textsc{python} package, \textsc{uncertainties} (\url{http://pythonhosted.org/uncertainties/}).

\bibliography{bibliography}

\appendix

\section{Propagation of Uncertainty}\label{App1}

Here, we provide formulae necessary to calculate uncertainties on properties of the disk and total galaxy. For the complementary equations for properties of the spheroid, see Equations~(7) and (10) from \citetalias{Davis:2018}.

\begin{equation}
\mathfrak{m_{\rm disk}}=-2.5\log\left(10^{-0.4\mathfrak{m_{\rm tot}}}-10^{-0.4\mathfrak{m_{\rm sph}}}\right )
\label{disk_app_mag}
\end{equation}

\begin{equation}
\delta\mathfrak{m_{\rm disk}}=\frac{\sqrt{(L_{\rm tot}\delta\mathfrak{m_{\rm tot}})^2+(L_{\rm sph}\delta\mathfrak{m_{\rm sph}})^2}}{L_{\rm tot}-L_{\rm sph}}
\label{disk_app_mag_error}
\end{equation}

\begin{equation}
\delta\mathfrak{M_{\rm tot}}=\sqrt{\delta\mathfrak{m_{\rm tot}}^2+\left[\frac{5(\delta d_L)}{d_L\ln(10)}\right]^2}
\label{total_abs_mag_error}
\end{equation}

\begin{equation}
\delta\mathfrak{M_{\rm disk}}=\sqrt{\frac{(L_{\rm tot}\delta\mathfrak{m_{\rm tot}})^2+(L_{\rm sph}\delta\mathfrak{m_{\rm sph}})^2}{(L_{\rm tot}-L_{\rm sph})^2}+\left[\frac{5(\delta d_L)}{d_L\ln(10)}\right]^2}
\label{disk_abs_mag_error}
\end{equation}

\begin{equation}
\delta \log{M_{*,\rm tot}}=\sqrt{\left(\frac{\delta\mathfrak{m_{\rm tot}}}{2.5}\right)^2+\left[\frac{2(\delta d_L)}{d_L\ln(10)}\right]^2+\left[\frac{\delta\Upsilon_*}{\Upsilon_*\ln(10)}\right]^2}
\label{total_mass_error}
\end{equation}

\begin{equation}
\delta \log{M_{*,\rm disk}}=\sqrt{\frac{(L_{\rm tot}\delta\mathfrak{m_{\rm tot}})^2+(L_{\rm sph}\delta\mathfrak{m_{\rm sph}})^2}{\left[2.5(L_{\rm tot}-L_{\rm sph})\right]^2}+\left[\frac{2(\delta d_L)}{d_L\ln(10)}\right]^2+\left[\frac{\delta\Upsilon_*}{\Upsilon_*\ln(10)}\right]^2}
\label{disk_mass_error}
\end{equation}

\section{Bayesian Prior and Posterior Values}\label{App2}

Here, we summarize the results of fitting our Bayesian models against the observational data sets of the $M_{\rm BH}$--$M_{\rm *,tot}$ (Table~\ref{table:priors_tot}) and $M_{\rm BH}$--$M_{\rm *,disk}$ (Table~\ref{table:priors_disk}) relations. In particular, we report the estimated quantiles at 2.5\%, 16\%, 50\%, 84\%, and 97.5\% for each parameter; from these can be read the median, 68\% (``$\pm1$\,$\sigma$''), and 95\% (``$\pm2$\,$\sigma$'') credible intervals. Illustrations of our fits are also presented in Figures \ref{Ewan_M_BH-M_tot_plot} and \ref{Ewan_M_BH-M_disk_plot}. From inspection of Tables~\ref{table:priors_tot} and \ref{table:priors_disk}, it is evident that our priors are strongly updated by the data.

\begin{deluxetable}{lrrrrr|rrrrr}
\tabletypesize{\normalsize}
\tablecolumns{11}
\tablecaption{Fitting Results of Our Model against the Observational Data Set $\left(\log{M_{\rm *,tot}},\, \log{M_{\rm BH}}\right)$\label{table:priors_tot}}
\tablehead{
\colhead{} & \multicolumn{5}{c}{Prior} & \multicolumn{5}{c}{Posterior}
}
\startdata
Quantile & 2.5\% & 16\% & 50\% & 84\% & 97.5\% & 2.5\% & 16\% & 50\% & 84\% & 97.5\% \\
\hline
\textit{Symmetric} slope & 0.03 & 0.19 & 1.00 & 5.30 & 39.35 & 2.12 & 2.56 & 3.05 & 3.62 & 4.14 \\
\textit{Conditional} $(Y|X)$ slope & 0.02 & 0.17 & 0.90 & 4.77 & 36.12 & 1.28 & 1.62 & 2.03 & 2.47 & 2.96 \\
\textit{Symmetric} $M_{\rm BH}$ scatter (dex)& 0.01 & 0.05 & 0.22 & 0.68 & 1.65 & 0.63 & 0.70 & 0.79 & 0.90 & 1.00 \\
\textit{Conditional} $M_{\rm BH}$ scatter (dex) & 0.01 & 0.05 & 0.21 & 0.67 & 1.61 & 0.45 & 0.51 & 0.58 & 0.66 & 0.75 \\
Normalized $X$-intercept, $X_0$ & 6.58 & 8.51 & 10.50 & 12.49 & 14.42 & 10.71 & 10.76 & 10.80 & 10.85 & 10.90 \\
Normalized $Y$-intercept, $Y_0$ & 3.08 & 5.01 & 7.00 & 8.99 & 10.92 & 6.98 & 7.11 & 7.25 & 7.38 & 7.52 \\
\enddata
\end{deluxetable}

\begin{deluxetable}{lrrrrr|rrrrr}
\tabletypesize{\normalsize}
\tablecolumns{11}
\tablecaption{Fitting Results of Our Model against the Observational Data Set $\left(\log{M_{\rm *,disk}},\ \log{M_{\rm BH}}\right)$\label{table:priors_disk}}
\tablehead{
\colhead{} & \multicolumn{5}{c}{Prior} & \multicolumn{5}{c}{Posterior}
}
\startdata
Quantile & 2.5\% & 16\% & 50\% & 84\% & 97.5\% & 2.5\% & 16\% & 50\% & 84\% & 97.5\% \\
\hline
\textit{Symmetric} slope & 0.03 & 0.19 & 0.99 & 5.27 & 39.79 & 2.08 & 2.41 & 2.83 & 3.38 & 4.04 \\
\textit{Conditional} $(Y|X)$ slope & 0.02 & 0.17 & 0.92 & 4.74 & 34.78 & 1.12 & 1.39 & 1.74 & 2.17 & 2.63 \\
\textit{Symmetric} $M_{\rm BH}$ scatter (dex) & 0.01 & 0.05 & 0.21 & 0.68 & 1.56 & 0.64 & 0.71 & 0.80 & 0.90 & 1.04 \\
\textit{Conditional} $M_{\rm BH}$ scatter (dex) & 0.01 & 0.05 & 0.22 & 0.68 & 1.57 & 0.49 & 0.55 & 0.62 & 0.70 & 0.81 \\
Normalized $X$-intercept, $X_0$ & 6.58 & 8.51 & 10.50 & 12.49 & 14.42 & 10.60 & 10.65 & 10.70 & 10.75 & 10.80 \\
Normalized $Y$-intercept, $Y_0$ & 3.08 & 5.01 & 7.00 & 8.99 & 10.92 & 6.98 & 7.11 & 7.24 & 7.37 & 7.50 \\
\enddata
\end{deluxetable}

\end{document}